\documentclass[reprint,amsmath,amssymb,pre]{revtex4-1}
\pdfoutput=1
\usepackage{graphicx}
\usepackage{dcolumn}
\usepackage{bm}

\begin{document}

\title{Minimal stochastic field equations for one-dimensional flocking}

\author{E. \'{O} Laighl\'{e}is}
\author{M. R. Evans}
\author{R. A. Blythe}
\affiliation{SUPA, School of Physics and Astronomy, University of Edinburgh, Peter Guthrie Tait Road, Edinburgh EH9 3FD, United Kingdom}

\date{\today}

\begin{abstract}
We consider the collective behaviour of active particles that locally align with their neighbours. Agent-based simulation models have previously shown that in one dimension, these particles can form into a flock that maintains its stability by stochastically alternating its direction. Until now, this behaviour has been seen in models based on continuum field equations only by appealing to long-range interactions that are not present in the simulation model. Here, we derive a set of stochastic field equations with local interactions that reproduces both qualitatively and quantitatively the behaviour of the agent-based model, including the alternating flock phase. A crucial component is a multiplicative noise term of the voter model type in the dynamics of the local polarization whose magnitude is inversely proportional to the local density. We show that there is an important subtlety in determining the physically appropriate noise, in that it depends on a careful choice of the field variables used to characterise the system. We further use the resulting equations to show that a nonlinear alignment interaction of at least cubic order is needed for flocking to arise. 
\end{abstract}

\maketitle

\section{\label{sec:Intro}Introduction}

The collective motion of self-propelled particles has been an active area of research for a number of years \cite{TonerTu2005_FlockingReview,Vicsek2012_CollectiveMotionReview,Ramaswamy2017_ActiveMatterReview}, with areas of particular interest including bird flocking \cite{Pearce2014_ProjectionBirdFlocks,LewisTurner2017_DensityDepthFlocks}, fish schooling patterns \cite{Katz2011_FishFlockingInteractions,Makris2009_FishExperimentalFlocking}, insect swarms \cite{Yates2009_LocustsAlternatingABM,Beekman2001_AntsCollectiveMotionTransition}, and the motion of bacteria \cite{Wu2009_BacteriaSwarmingAlternatingDirection}, among others. Interest in these systems lies in the fact that they are internally driven, and therefore generate complex non-equilibrium behaviours at the macroscopic scale. For example, in two dimensions, one can find long-range orientational order \cite{Vicsek1995_CollectiveMotionModel} in which the particles move in
a randomly selected direction thus breaking the continuous  rotational symmetry---such continuous symmetry breaking is precluded from
equilibrium models in two dimensions.There are two main methods to study such systems: agent-based models or continuum field equations. Simulations of stochastic agent-based models \cite{Vicsek1995_CollectiveMotionModel,Vicsek1999_1dFlocking,Chate2008_VicsekFirstOrderTransition} allow changes in behaviour to be explored as inter-particle interactions are varied, while a deeper theoretical understanding can be gained by deriving and analysing equations that govern the time evolution of particle density and velocity fields \cite{TonerTu1998_FieldModelFlocking,SolonTailleur2013_ActiveIsingFlockingTransition}. It is important to recognise that these approaches are complementary. Simulations allow fine-grained control over the microscopic dynamics of the system, but are typically limited in the range of system sizes and timescales that can be accessed. On the other hand, field equations are usually free from such restrictions, but instead necessitate some degree of approximation or coarse-graining to render them tractable.

Here, our focus is on the case of flocking in one spatial dimension (1d), as this exhibits some interesting behaviours that are not observed in higher dimensions, and for which a theoretical understanding through field equations is incomplete.  Of particular interest is the case where a flock---that is, a localised group of particles moving in the same direction---is maintained by {\em alternating} between the two possible directions that are available in 1d (i.e., left and right). Thus the order of the flock  is maintained, but as the timescale for the flock to alternate grows only logarithmically with system size, there is no spontaneous symmetry breaking. This effect was first studied in a simulation model \cite{OLoan1999_AlternatingFlock} in which interactions are local and involve some degree of stochasticity in the dynamics.  Various sets of field equations involving only local interactions have been constructed for active particles in 1d, both with \cite{Vicsek1999_1dFlocking,Vicsek2000_EarlyFlockingReview} and without noise \cite{SolonTailleur2013_ActiveIsingFlockingTransition,SolonChateTailleur2015_LiquidGasVicsek,Solon2015_IsingVicsekLiquidGas}. These demonstrate an instability for particles to form into a discrete number of bands or as a single large flock, but do not display an alternating behaviour as far as we are aware. However, both deterministic and stochastic equations with non-local interactions \cite{Eftimie2007_FlockingPatterns,Eftimie2013_CollectiveBehaviour,Dyson2015_LocustFlockingFieldModel} \emph{do} generate a variety of complicated patterns, including the alternating state (``zig-zags'') of interest here, and it has been suggested that non-local interactions are necessary to reproduce such behaviours in the deterministic setting \cite{Eftimie2012_FlockingModelsReview}. Thus the question of whether it is possible to represent the alternating state in a system of local \emph{stochastic}  field equations is left open.

In this work, we return to the agent-based model introduced in \cite{OLoan1999_AlternatingFlock} that is defined on a lattice and use this to construct a set of stochastic field equations in one spatial dimension. Through numerical integration of these equations, we show that they reproduce the alternating behaviour seen in the agent-based model both qualitatively and quantitatively. The strategy we adopt to obtain the equations comprises two steps. First, we perform a Kramers-Moyal expansion \cite{Gardiner2009_HandbookStochasticMethods} in the particle density (the fraction of particles at a given lattice site) and polarization (the fraction of particles at a site that are moving to the left subtracted from the corresponding fraction moving to the right), and truncate at second order. Then, we introduce a continuum approximation to the lattice-based model, which gives rise to a set of equations that involve spatial derivatives of the two fields, and a noise term. Although these are both widely-used procedures, we identify an important subtlety that arises: the choice of fields used to describe the system is crucial. In the case where we describe the difference in densities of the right- and left-moving particles in terms of the polarization, which is normalised by the local particle density and therefore constrained to lie between $-1$ and $1$, we obtain a multiplicative noise that is of the voter type to lowest order \cite{Dickman1995_VoterTypeNoise,Dornic2001_UniversalityClassVoterModel,RussellBlythe2011_NoiseInducedTransition} and find that this generates the fluctuations that are required to precipitate the wholesale reversal of a flock. By contrast, if we use instead the local momentum (i.e., the difference in densities of the left- and right-moving particles), we obtain a noise that is additive to leading order and does not display the alternating behaviour. The reason for this is that by truncating the Kramers-Moyal expansion at second order is an approximation that can discard relevant contributions under certain choices of variables, but not others. 

Having established a set of equations that reproduces the correct macroscopic physics, we then use them to identify the leading order terms in the inter-particle interactions that are required, and therewith a minimal model for one-dimensional flocking. In addition to voter-type noise, we find that it is necessary that the propensity for particles to align with their neighbours is a nonlinear (specifically, at least cubic) function of local polarization. A linear alignment rule reduces to non-interacting run-and-tumble particles \cite{Schnitzer1993_RunAndTumble} that do not flock.

\section{\label{sec:Model}From an agent-based model to stochastic field equations}
\subsection{\label{sec:ABM}Agent-based model}

We begin by defining the agent-based model that serves as the starting point for deriving the stochastic field equations. This model is inspired by that introduced by O'Loan and Evans \cite{OLoan1999_AlternatingFlock}, as under certain conditions the particles form a cohesive flock that stochastically alternates direction.

\begin{figure}
    \centering
    \includegraphics[scale=0.24]{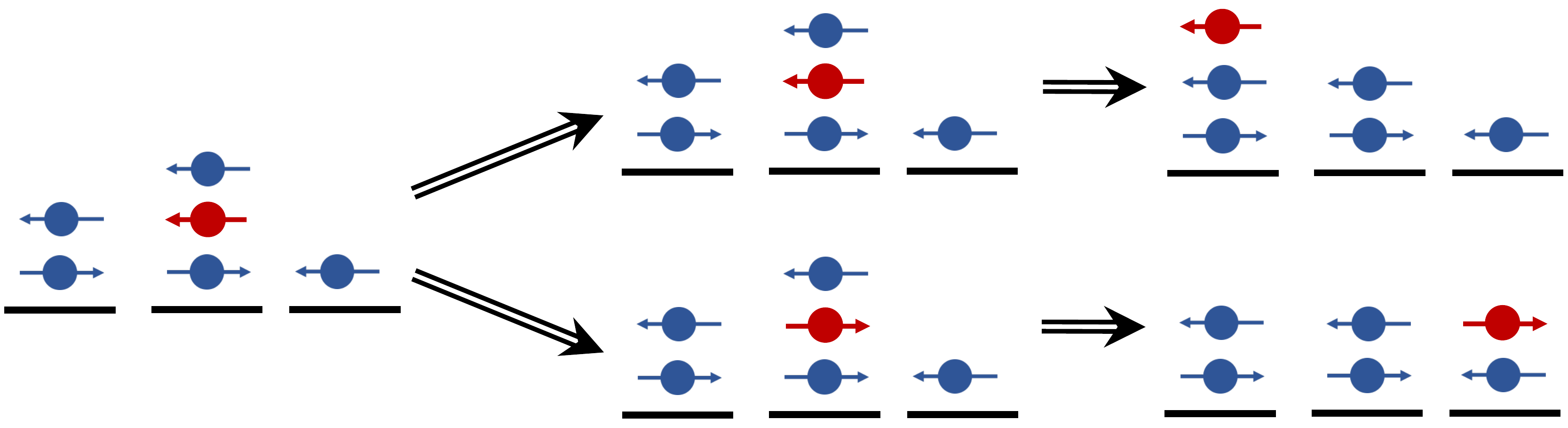}
    \caption{Possible updates for a chosen ABM (red - color online). With probability $W^-\left(-\frac{1}{3}\right)$, the particle retains its direction and moves one site to the left (top option). Alternatively, the particle switches direction and moves one site right with probability $W^+\left(-\frac{1}{3}\right)$ (bottom option)}
    \label{fig:ABM_Probs}
\end{figure}

In this model, $N$ identical particles are distributed across $L$ lattice sites in a ring geometry, with no maximum occupancy at each site. Each particle is initially assigned a random direction $v^{\alpha} = \pm 1$. In a single update, a particle $\alpha \in \left\lbrace 1,\dots,N \right\rbrace$ at site $x^{\alpha} \in \left\lbrace 1,\dots,L \right\rbrace$ is chosen at random. It then hops either to the left ($-$) or the right ($+$) (updating its velocity to $-1$ or $+1$ respectively) with rate $W^{\pm}(\theta(x^{\alpha}))$ defined as 
\begin{equation}
    \label{eq:rates}
    W^{\pm}(\theta) = \frac{1}{2}\left[1 \pm g(\theta)\right]
\end{equation}
as shown in Fig.~\ref{fig:ABM_Probs}. Here, the \emph{polarization} $\theta(x)$ is defined as
\begin{equation}
    \label{eq:theta}
    \theta(x) = \dfrac{\sum_{\alpha=1}^N v^{\alpha} \delta_{x,x^{\alpha}}} {\sum_{\alpha=1}^N \delta_{x,x^{\alpha}}}
\end{equation}
and lies in the range $\theta(x) \in [-1,1]$. Concretely, the polarization is equal to the fraction of particles \emph{at a site} that are moving to the right, minus the corresponding fraction moving to the left.

The difference in the probabilities $W^{+}(\theta) - W^{-}(\theta) = g(\theta)$ specifies the velocity (normalised to the range $[-1,1]$) with which particles at position $x$ move if the polarization $\theta(x) \in [-1,1]$. We can therefore think of $g(\theta)$ as specifying the extent to which particles align themselves with the local polarization (i.e., the velocity of the other particles). In the original model of \cite{OLoan1999_AlternatingFlock}, particles aligned themselves with the majority with probability $(1-\eta)$, and in the opposite direction with probability $\eta$, which corresponds to the choice $g(\theta) = (1-2\eta)  {\rm sgn}(\theta)$.  Here, we adopt a form that regularises the step at $\theta=0$, namely
\begin{equation}
\label{eq:gdef}
g(\theta) = (1-2\eta)\frac{\tanh(\beta \theta)}{\tanh(\beta)} \;.
\end{equation} 
With this form, we recover the majority alignment rule in the limit $\beta\to\infty$, and a linear alignment rule as $\beta\to0$. We will later show that to achieve the alternating flock behaviour, it is necessary for $g(\theta)$ to be nonlinear.  One further difference from the  model of \cite{OLoan1999_AlternatingFlock} is that in that work, particles were also sensitive to the polarization on nearest-neighbour sites. We have found that this is not in fact required for the alternating state to emerge, and have thus excluded such interactions for simplicity.

\begin{figure}
    \centering
    \includegraphics[scale=0.28]{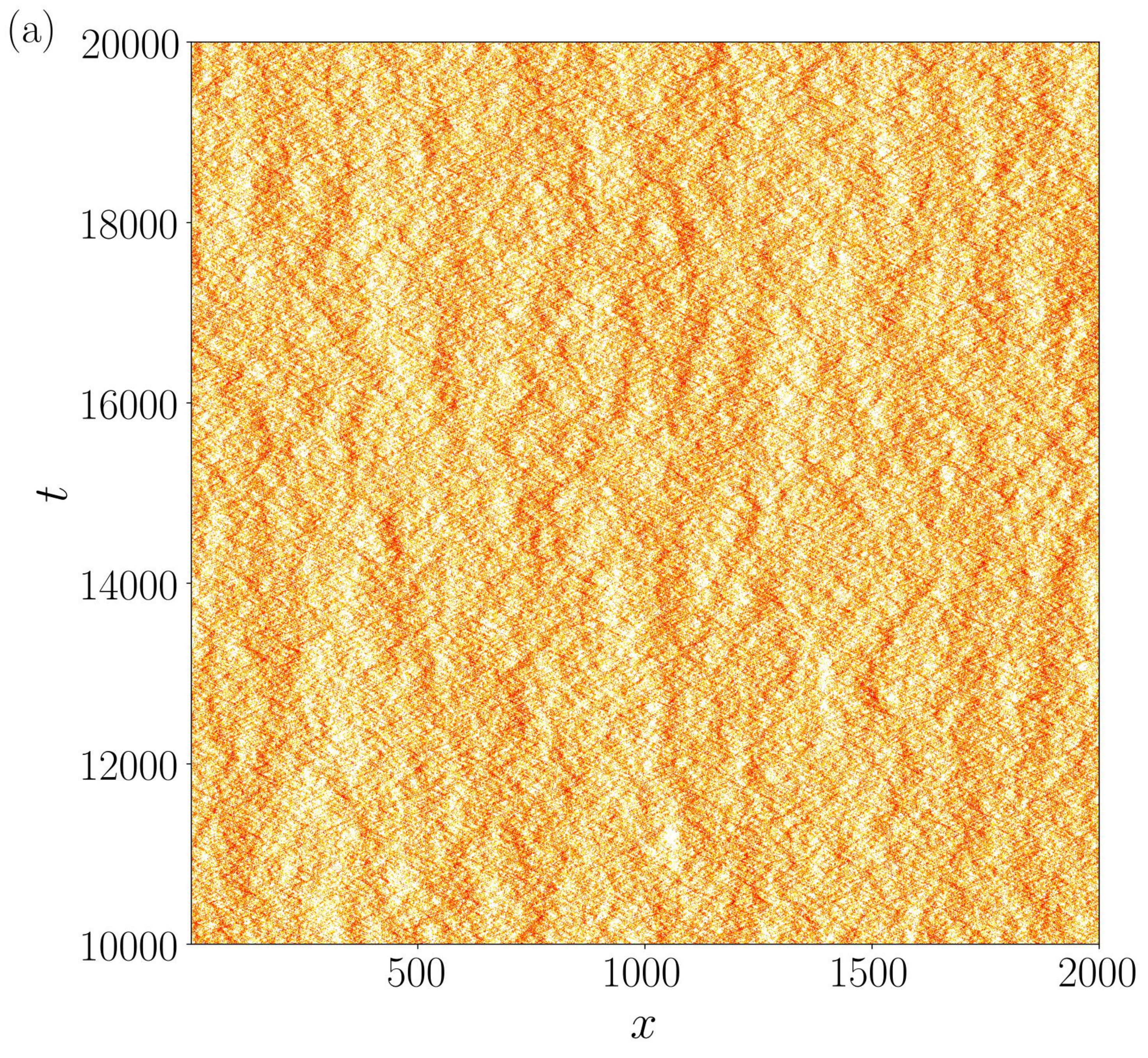}\\
    \includegraphics[scale=0.28]{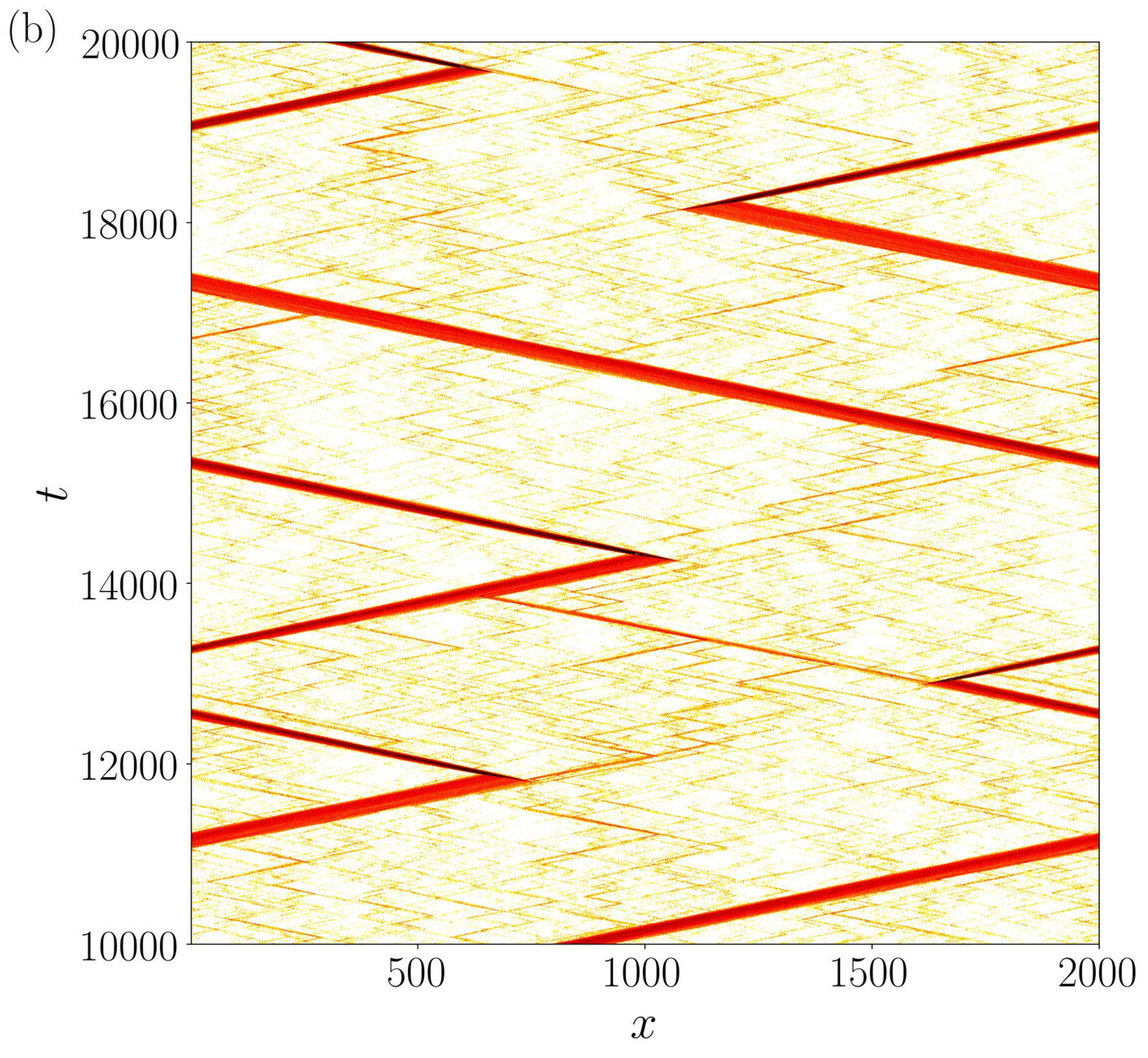}
    
    \caption{Sample space-time trajectories of the fractional density $\rho$ with periodic boundaries displaying the two behaviours observed in the agent-based model (darker regions indicate higher density). a) Paramagnetic phase ($\eta = 0.2$). b.) Alternating flock ($\eta = 0.02$). We have $N = L = 2000$ and $\beta = 2.0$ for both.}
    \label{fig:ABM-Phases}
\end{figure}

\begin{figure}
    \centering
    \includegraphics[width=\linewidth]{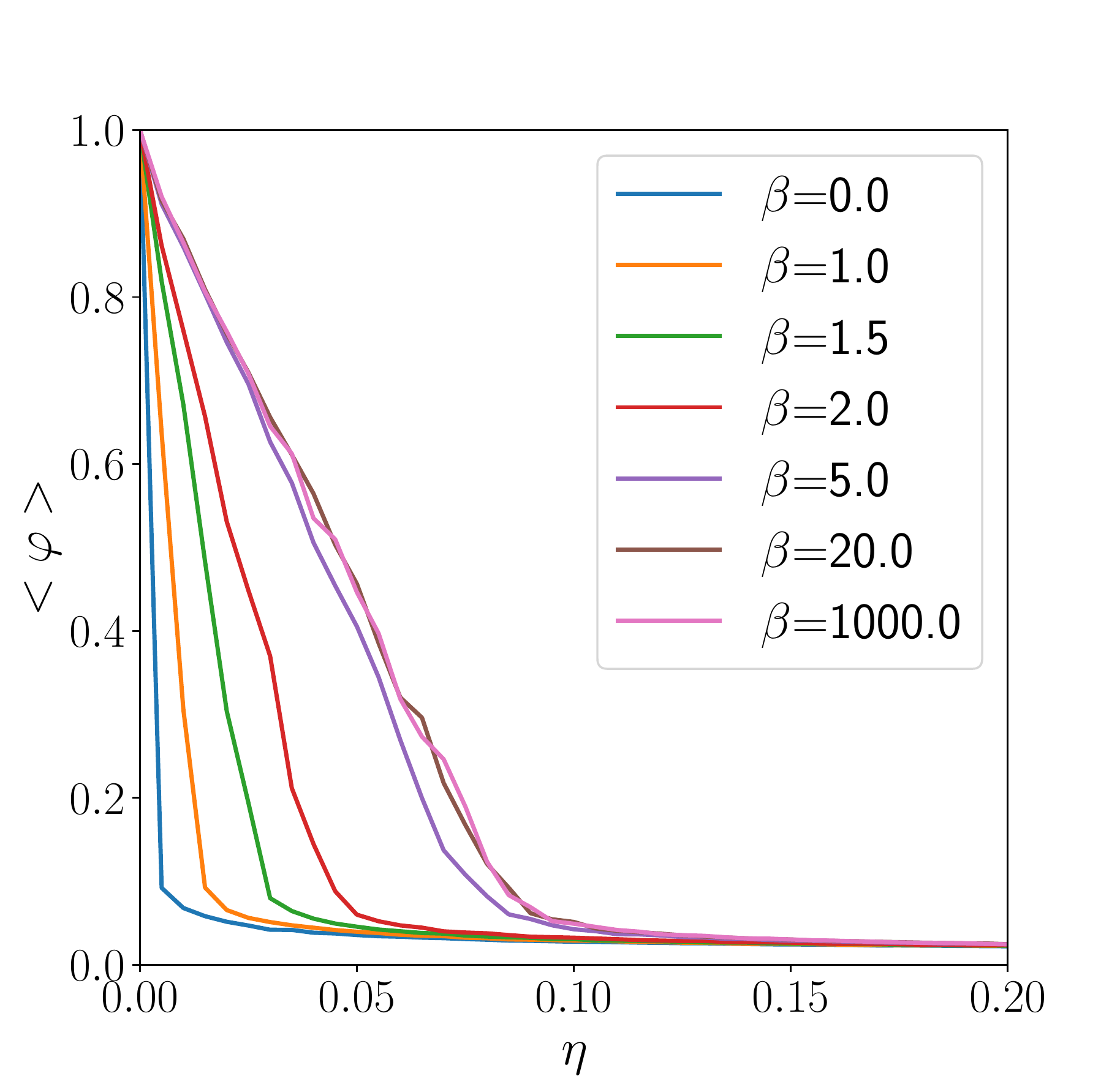}
    
    \caption{Order parameter (\ref{eq:order}) as a function of the anti-alignment rate $\eta$ for various degrees of nonlinearity $\beta$ in the alignment interaction (\ref{eq:gdef}). When the anti-alignment is sufficiently small, the system orders into a state where particles tend to move in the same direction (i.e., flock).}
    \label{fig:ChangingBeta}
\end{figure}

Fig.~\ref{fig:ABM-Phases} demonstrates that the basic phase behaviour of this model is the same as that established in \cite{OLoan1999_AlternatingFlock}. Depending on the values of $\beta$ and $\eta$ above, the system will manifest one of two phases: a disordered paramagnetic phase with uniform average density and negligible average on-site polarization; or an ordered phase where a region of larger density propagates through the ring, alternating direction stochastically. In Fig.~\ref{fig:ChangingBeta} we plot the global velocity 
\begin{equation}
\label{eq:order}
\varphi = \frac{1}{N} \left| \sum_{\alpha=1}^{N} v^{\alpha} \right|
\end{equation}
against $\eta$. This locates  the transition between these two states for different values of $\beta$, with a non-zero value of $\varphi$ in the ordered (alternating) state at low $\eta$.

A third phase is seen in both the low noise ($\eta \rightarrow 0$) and the high-density ($N \rightarrow \infty$ with finite $L$) limits. Here, the particles spread out across the lattice uniformly travelling in a single direction and we term it the ``ferromagnetic'' phase. However, Ramaswamy \cite{Ramaswamy2010_ActiveMatterReview} argues that this is not a true thermodynamic phase, as it cannot exist if $\eta$ is nonzero and $N/L$ is finite in the $L \rightarrow \infty$ limit. This is due to an instability to transition to the alternating state, where the symmetry-breaking property of the ferromagnetic state is no longer observed.

Observations made in \cite{OLoan1999_AlternatingFlock} for an agent-based model and \cite{Raymond2006_Flocking} for a related mean-field model without alternating behaviours show that the flock in the alternating phase diffuses slowly between directional flips, with these flips appearing to serve the purpose of condensing the flock and keeping it together. If the time between successive flips is too long, this can result in a temporary transition to the ferromagnetic pseudo-phase until a fluctuation causes the flock to return.

\subsection{\label{sec:SPDE}Stochastic field equations}

We now show how to construct a set of field equations that correctly reproduces the behaviour described above. The first step is to employ a  Kramers-Moyal expansion \cite{Kramers1940_FPDerivation,Moyal1949_FPDerivation,Gardiner2009_HandbookStochasticMethods}. If this is truncated at second order, we arrive at an approximation for the time evolution of a stochastic variable $\psi$ of the form
\begin{equation}
    \label{eq:KramersMoyal}
    \frac{d\psi}{dt} \approx \frac{\left< \delta \psi \right>}{\delta t} + \sqrt{\frac{\left< \delta \psi^2 \right>}{\delta t}} \xi(t) 
\end{equation}
where $\delta \psi$ represents the possible changes in $\psi$ in a single timestep $\delta t$ and $\xi(t) \sim \mathcal{N}(0,1)$ is a white Gaussian noise. This equation is to be interpreted in the It\^o sense. For our system, the stochastic variables of interest are the local fractional density, \begin{equation}
\rho(x,t) = \frac{1}{N}\sum_{\alpha = 1}^N \delta(x-x^{\alpha})\;,
\end{equation}
and the local polarization, $\theta$, as in (\ref{eq:theta}). In order to view these variables as continuous, we must consider the large $N$ limit of our system. As we shall explore later the choice of stochastic variables $\psi$ matters here, as a consequence of truncating the expansion at second order to arrive at an It\^o SDE (for general discussion of It\^o versus Stratonovich noise, see \cite{Gardiner2009_HandbookStochasticMethods}).

Given the rules for the agent-based model set out in Sec.~\ref{sec:ABM} above, we can determine the changes in the fractional density and polarization at site position $x_i$ arising from a single update as follows:
\begin{align}
    \label{eq:rhoDiscrete}
    \delta \rho_i(t) &= \left\lbrace \begin{array}{cl}
    \frac{1}{N} & \mbox{with prob. } W_{i-1}^+ \rho_{i-1} + W_{i+1}^- \rho_{i+1} \\
    - \frac{1}{N} & \mbox{with prob. } \rho_i \\
    0 & \mbox{otherwise}
\end{array} \right. \\
    \label{eq:thetaDiscrete}
    \delta \theta_i(t) &= \left\lbrace \begin{array}{cl}
    \frac{1 - \theta_i}{N \rho_i} & \mbox{with prob. } W_{i-1}^+ \rho_{i-1}\\
    - \frac{1 + \theta_i}{N \rho_i} & \mbox{with prob. } W_{i+1}^- \rho_{i+1} \\
    \frac{1 + \theta_i}{N \rho_i} & \mbox{with prob. } \rho_i^- \\
    - \frac{1 - \theta_i}{N \rho_i} & \mbox{with prob. } \rho_i^+ \\
    0 & \mbox{otherwise}
\end{array} \right.
\end{align}
where $\rho_i^+$ and $\rho_i^-$ represent the local fractional densities of right- and left-moving particles respectively. The second step of the derivation is to take the limit of large lattice size, $L \rightarrow \infty$, which is achieved by making the replacements $\rho_i \to \rho(x)$ and $\theta_i \to \theta(x)$, with $x=ia$, and Taylor expanding expressions like $\rho(x\pm a)$ to second order in $a$. Finally, the lattice spacing $a$ is set to $1$. With the assumption that the density and polarization fields are uncorrelated in space, this procedure yields the expressions
\begin{align}
    \label{eq:KMRho1}
    \left<\delta \rho \right> &= \dfrac{1}{N} \left[-\frac{\partial \left(g(\theta) \rho\right)}{\partial x} + \dfrac{1}{2} \dfrac{\partial^2 \rho}{\partial x^2}\right] \\
    \label{eq:KMRho2}
    \left<\delta \rho^2 \right> &= \dfrac{1}{N^2} \left[2\rho -\frac{\partial \left(g(\theta) \rho\right)}{\partial x} + \dfrac{1}{2} \dfrac{\partial^2 \rho}{\partial x^2}\right] \\
    \label{eq:KMTheta1}
    \left<\delta \theta \right> &= \dfrac{1}{N} \left[g(\theta) - \dfrac{\partial \ln{\rho}}{\partial x} + \dfrac{1}{2\rho} \dfrac{\partial^2 (g(\theta) \rho)}{\partial x^2} \right] - \dfrac{\theta}{\rho} \left<\delta \rho \right> \\
    \label{eq:KMTheta2}
    \left<\delta \theta^2 \right> &= \dfrac{2}{N^2\rho} \left[1 - \theta g(\theta)\right] \;.
\end{align}

Inserting the above into (\ref{eq:KramersMoyal}), setting $\delta t = N^{-1}$, and keeping terms of $O\left(1\right)$ gives the following coupled stochastic differential equations for the time evolution of the fractional density and polarization fields
\begin{align}
    \label{eq:LangevinRho}
    \dfrac{\partial \rho}{\partial t} &= -\frac{\partial \left(g(\theta) \rho\right)}{\partial x} + \dfrac{1}{2} \dfrac{\partial^2 \rho}{\partial x^2} \\
    \label{eq:LangevinTheta}
    \dfrac{\partial \theta}{\partial t} &= g(\theta) - \theta - \dfrac{\partial \ln{\rho}}{\partial x} + \dfrac{1}{2\rho} \dfrac{\partial^2 (g(\theta) \rho)}{\partial x^2} - \theta \dfrac{\partial \ln{\rho}}{\partial t}\nonumber\\
    & \qquad + \sqrt{\dfrac{2}{N\rho}}\sqrt{1 - \theta g(\theta)}\ \xi(x,t) \;.
\end{align}
Equations (\ref{eq:LangevinRho}, \ref{eq:LangevinTheta}) form the basis for our field equation description of alternating flocking.

First  we must comment on why we have dropped noise term in (\ref{eq:LangevinRho}) but not in (\ref{eq:LangevinTheta}), as na\"{\i}vely both are of the same order, $1/N^2$, in (\ref{eq:KMRho2}) and (\ref{eq:KMTheta2}), respectively. The key difference is that the magnitude of the second jump moment $\langle\delta\theta^2\rangle$ is inversely proportional to the density, $\rho$, which can be of order $1/N$, particularly in the nose and the tail of a flock. Consequently in these regions, the stochastic and deterministic terms are of similar magnitude, and as argued in \cite{OLoan1999_AlternatingFlock}, it is these fluctuations that are responsible for reversing the direction of the flock. This we will confirm explicitly in the following. A further implication of dropping the noise term in (\ref{eq:LangevinRho}) is that the equation has the form of a continuity equation and respects the fact that the density $\rho$ is a conserved quantity across the system. If we had retained the second order term (\ref{eq:KMRho2}) in (\ref{eq:LangevinRho}), the resulting equation would no longer conserve density. This would seem to arise from the assumption that density and polarization fields are spatially uncorrelated. If we were to retain these correlations, we would end up with a nontrivial noise term arising whose amplitude is given by the Cholesky decomposition of the correlation matrix \cite{Pavliotis2014_StochasticProcesses,Dereniowski2004_CholeskyDecomposition}. In principle, this should should lead to a conservative noise, as obtained in other methods of deriving field equations from microscopic interactions such as those described by Bertin \cite{Bertin2013_ActiveNematics} and Dean \cite{Dean1996_ConservedNoise}. However, as our aim is to describe a simple model for this behaviour in 1d, we have chosen instead to drop this noise, as it is of lower order in $N$ than the terms we have retained. This is justified a posteriori by the good quantitative agreement between numerical integration of these equations and the agent-based simulations that we report below.

\subsection{\label{sec:Beta}Numerical integration of the stochastic field equations}

To investigate the behaviour of the system described by the stochastic field equations (\ref{eq:LangevinRho}) and (\ref{eq:LangevinTheta}), we numerically integrate them. In the deterministic terms, we follow the standard approach of replacing space and time derivatives with finite differences: throughout this work we use $\delta x= 1$ and $\delta t= 0.01$. The standard (Euler-Maruyama) method for handling the stochastic term, wherein one replaces the combination $\xi(x,t)\sqrt{\delta t}$ with a Gaussian random variable with zero mean and variance $\delta t$ is not well adapted to the problem at hand. The issue is that as $\theta\to\pm1$, this can generate unphysical values of the polarization (i.e., $|\theta|>1$).

To handle this difficulty we first linearise the function $g(\theta)$ in the noise term whilst leaving it as a general function in the deterministic terms. That is, we put $g(\theta)=\theta$ under the square root in (\ref{eq:LangevinTheta}). This then yields a noise term of the voter type \cite{Dickman1995_VoterTypeNoise,Dornic2001_UniversalityClassVoterModel,RussellBlythe2011_NoiseInducedTransition} (also characteristic of the Wright-Fisher model \cite{Crow1970_WrightFisher,Blythe2007_ExactSolnKolmogorovGeneticDrift}), causing the noise to vanish at the boundary points $\theta=\pm1$. As we are trying to find a minimal model for 1d flocking, we need to keep only the necessary terms for the alternating flock to be observed. A crucial feature of the dynamics is that $\theta$ must be constrained to the interval $-1 \le \theta \le 1$:  the noise term in (\ref{eq:LangevinTheta}) satisfies this requirement at the lowest order in powers of $\theta$ and $\eta$. Therefore we may neglect higher order terms in $\theta$ and $\eta$ here. Numerically, however, it is still possible that a Gaussian random number large enough to exit the boundaries is generated. In the following we compare results using two strategies to deal with this. The first, and more straightforward strategy, is simply to truncate $\theta$ to the physical range whenever it is left. The more sophisticated strategy, inspired by the approach of \cite{Michaud2017_WeakNoiseKramersMoyalModification}, is to replace the Gaussian random variable in the numerical integration scheme with a variable drawn from a distribution that is defined only over the physical range, and has the required mean and variance.

More precisely, this second approach involves integrating the stochastic and deterministic parts of the equation separately using Hamiltonian operator splitting on the Fokker-Planck equation equivalent to (\ref{eq:LangevinTheta}) \cite{MoroSchurz2007_NIOperatorSplitting}. This involves sampling an intermediate value of $\theta$, $\theta_{\rm int}$, from the distribution obtained by solving $\partial_t P(\rho,\theta,t) = \left(N\rho\right)^{-1} \partial_{\theta\theta}\left[(1-\theta^2)P\right]$. By transforming to another variable $y = \frac{1}{2} (1+\theta)$, this distribution can be transformed to one with a known solution \cite{Kimura1955_WrightFisherFPSolution}. However, unlike other forms of noise with a known distribution such as directed percolation \cite{Dornic2005_MultiplicativeNoise}, this distribution cannot be written in a form that allows it to be easily sampled.

However, it can be shown that, given an initial condition $\theta$, the mean of $y$ after a time $\delta t$ is equal to $\mu = \frac{1}{2}\left( 1+\theta \right)$ and has a variance 
\begin{equation}
\sigma^2 = (1-\theta^2)\left(1-\exp\left[-\frac{2}{N\rho}\delta t\right]\right) \;.
\end{equation}
An appropriate distribution to sample $y$ from is the Beta distribution, $P(y) \propto y^{\alpha} (1-y)^{\beta}$ for $y \in [0,1]$, with the parameters $\alpha$ and $\beta$ chosen so that the distribution has the desired mean and variance. This is the natural distribution to choose because it is defined only for a finite interval, and has an independent mean and variance. Since we have already truncated the Kramers-Moyal expansion at second order to obtain the stochastic field equations (\ref{eq:LangevinRho}) and (\ref{eq:LangevinTheta}), we do not expect differences between the Beta and Gaussian distributions in the third and higher moments to be physically relevant for small timesteps.  Once the intermediate value $\theta_{\rm int}=2y-1$ is sampled, we then integrate the deterministic part of (\ref{eq:LangevinTheta}) numerically using the usual Euler method.

\section{\label{sec:Results}Minimal model for 1d flocking}

We now show that the stochastic field equations (\ref{eq:LangevinRho}) and (\ref{eq:LangevinTheta}) generate realisations of the density and polarization fields that are in both qualitative and quantitative agreement with the agent-based model. We also show that a nonlinear function $g(\theta)$ is required in the deterministic part of the equations in order to replicate  alternating flocks.

\subsection{Alternating flocks}

\begin{figure}[t]
    \centering
    
	\includegraphics[scale=0.26]{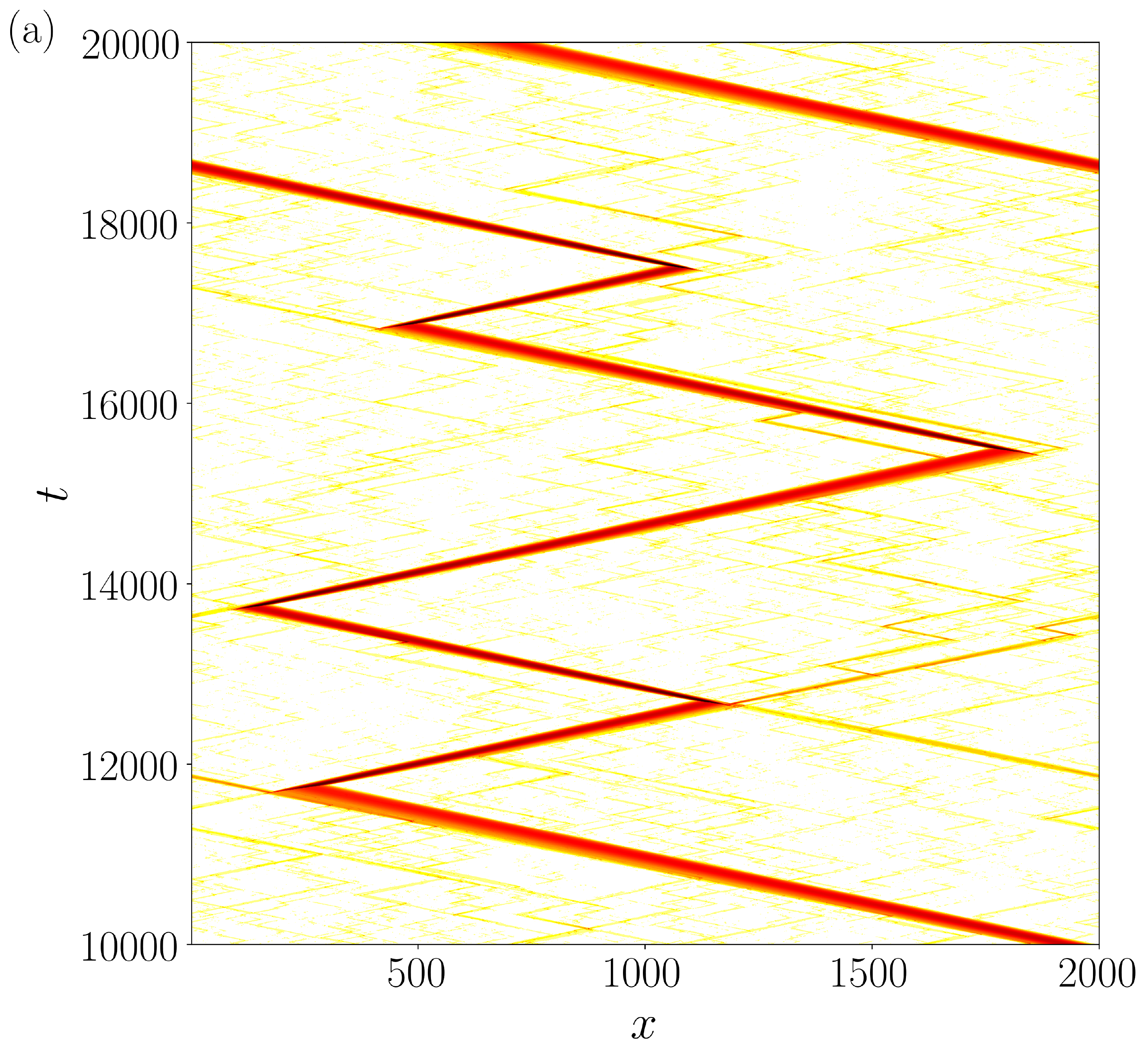}\\
	\includegraphics[scale=0.26]{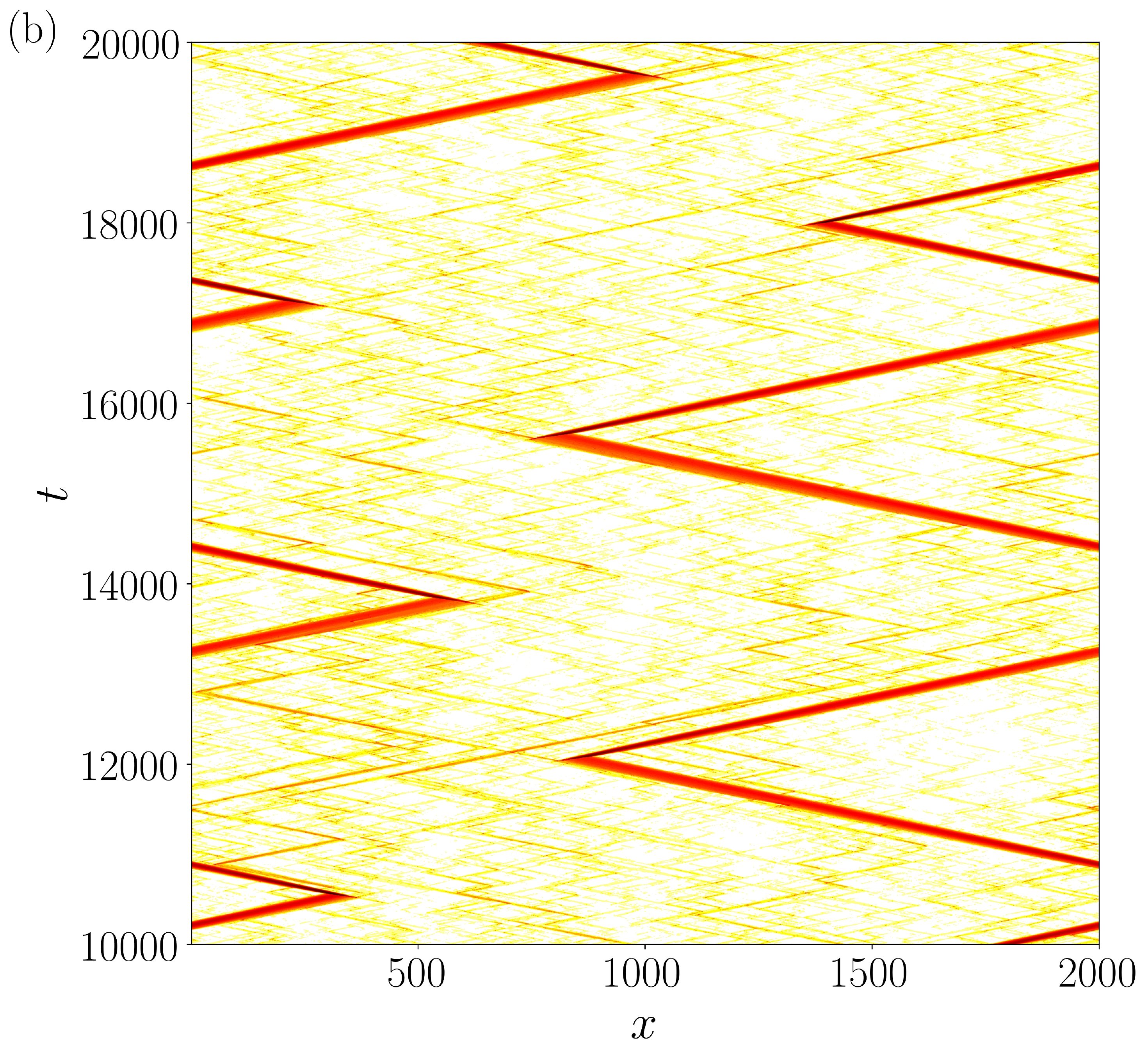}
    
    \caption{Numerically integrated solution of (\ref{eq:LangevinRho}) using (a) truncated Gaussian and (b) Beta distributed noise. While both display similar behaviours in terms of the main flock, the density of splinter flocks using Gaussian noise is lower than expected from the agent-based model. Both solutions have $N = L = 2000$, $\eta = 0.02$ and $\beta = 2.0$ and random initial starting conditions.}
    \label{fig:ThetaNoise}
\end{figure}

In the agent-based model we found that the alternating phase exists for small values of the anti-alignment parameter $\eta$. Specifically, in Fig.~\ref{fig:ABM-Phases}, the alternating state is shown for the case $\eta=0.02$ and with $\beta=2$ in the alignment interaction (\ref{eq:gdef}). In Fig.~\ref{fig:ThetaNoise} we show the result of integrating the stochastic field equations (\ref{eq:LangevinRho}) and (\ref{eq:LangevinTheta}) at the same point in parameter space, under both the truncated Gaussian and Beta sampling strategies.

In both cases we find that a flock forms and reverses its direction at random intervals, as in the agent-based model. Closer inspection of the data suggest that the flocks are more stable in the case of truncated Gaussian noise, with an apparently smaller number `splinter' flocks being emitted from the main flock than in either the agent-based model or when the Beta sampling strategy is used. 

\begin{figure}
    \centering
	\includegraphics[scale=0.40]{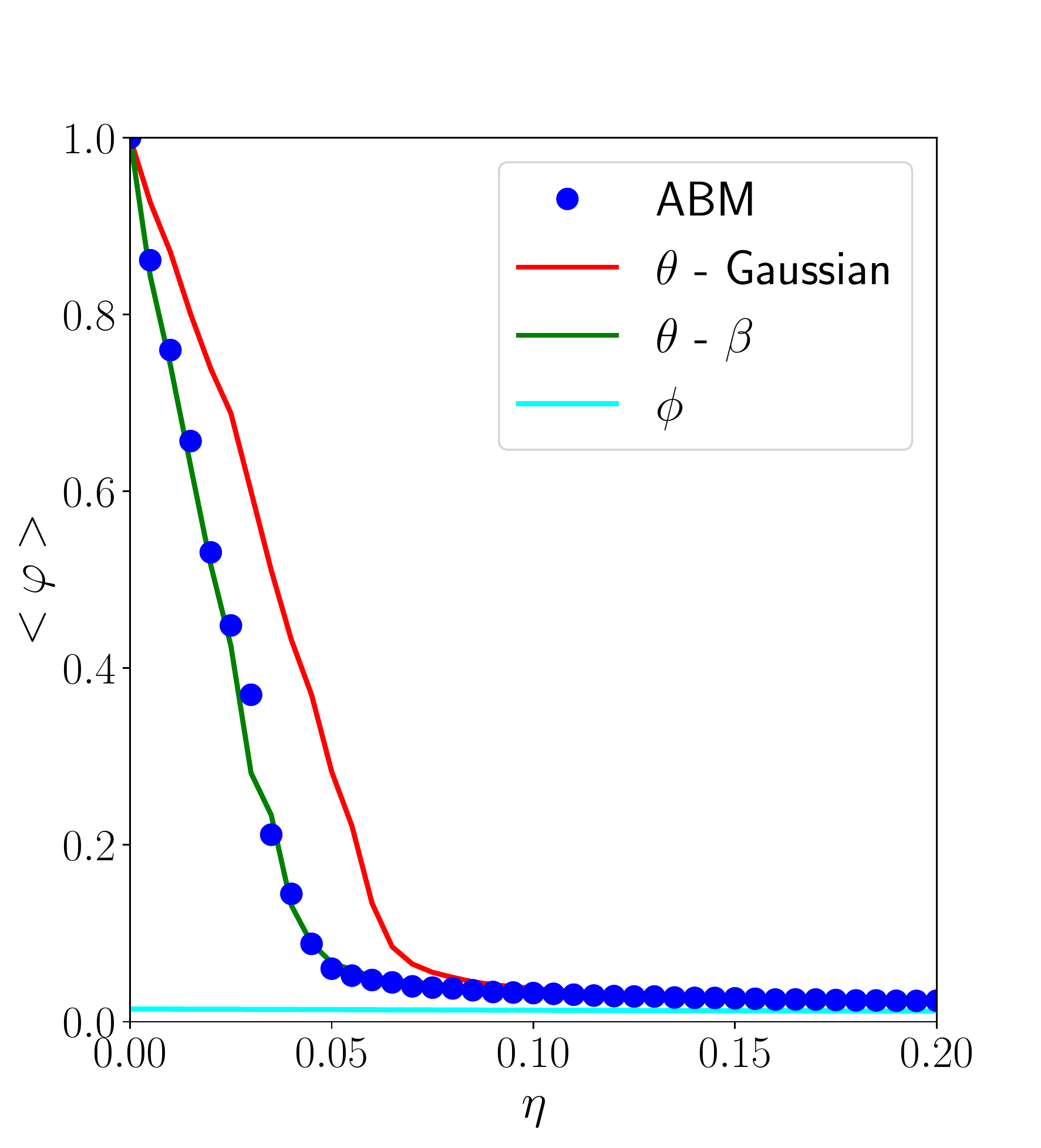}
    \caption{Plot of the change in the average order parameter $\left<\varphi\right>$ against anti-alignment strength $\eta$ with $N=L=2000$ and $\beta = 2.0$. The behaviour of the agent-based model (ABM) is described by (\ref{eq:LangevinRho}) and (\ref{eq:LangevinTheta}) with $\beta$ distributed sampling ($\theta - \beta$, the green line (color online)). Using truncated Gaussian noise instead gives a higher than expected global velocity, while using the equations derived via the Kramers-Moyal expansion of the momentum $\phi$ yields a negligible global velocity for all parameter values. There is a transition between the alternating state and paramagnetic state at $\eta \approx 0.05$. The exact value depends on the choice of $\beta$ and of $g$.}
    \label{fig:OrdParamVsEta}
\end{figure}

To investigate these differences more systematically, we investigate the behaviour of the order parameter $\varphi$ defined by Eq.~(\ref{eq:order}) as a function of the parameter $\eta$, recalling that flocking is seen for small $\eta$ and a homogeneous phase is seen for large $\eta$. Fig.~\ref{fig:OrdParamVsEta} shows results from the agent-based model along with those obtained from the field equations using the two different noise sampling strategies. We see that there is a transition to flocking in both cases, but that the truncated Gaussian strategy leads to the order parameter being systematically overestimated (and therewith the critical $\eta$). On the other hand, the results using the Beta sampling strategy are in excellent agreement with those from the agent-based model, showing that (\ref{eq:LangevinRho},\ref{eq:LangevinTheta}) quantitatively capture the dynamics of the interacting particle system.

\subsection{\label{sec:VarChoice}Choice of field variables}

Starting from (\ref{eq:LangevinRho}) and (\ref{eq:LangevinTheta}) 
we can obtain an alternative set of field equations using 
a momentum-like variable
\begin{equation}
\phi = \rho\theta\;.
\end{equation}
The pair $(\rho, \phi)$ then contain the same information as $(\rho, \theta)$ and in principle ought to give an equivalent description.
We apply an It\^o transformation to the equations (\ref{eq:LangevinRho}) and (\ref{eq:LangevinTheta}). Taking $g(\theta)=\theta$ under the square root, as previously, yields
\begin{eqnarray}
    \label{eq:LangevinPhi}
    \dfrac{\partial \phi}{\partial t} &=& g'\left(\phi,\rho\right)\rho - \phi - \dfrac{\partial \rho}{\partial x} + \dfrac{1}{2} \dfrac{\partial^2 \left(g'\left(\phi,\rho\right) \rho\right)}{\partial x^2}\nonumber\\
    &&+ \sqrt{\dfrac{2}{N\rho}}\sqrt{\rho^2 - \phi^2}\ \xi(x,t)
\end{eqnarray}
where $g'(\phi,\rho) = g\left(\frac{\phi}{\rho}\right)$. Although, this would appear to introduce issues when $\rho = 0$,  these issues are already present with the original definition of $\theta$. Numerically integrating (\ref{eq:LangevinRho}) and (\ref{eq:LangevinPhi}) instead of (\ref{eq:LangevinRho}) and (\ref{eq:LangevinTheta}) changes none of the behaviour of the system.

We now show that difficulties occur if instead of using
(\ref{eq:LangevinTheta}) we try to derive  equations for
the $(\rho, \phi)$ system using a Kramers-Moyal expansion from the microscopic dynamics to second order.
We shall show that this approach can lead to equations that do {\em not} describe a system that orders as $\eta$ is reduced.  

When we follow through the Kramers-Moyal expansion, as described in Section~\ref{sec:SPDE}, but using $(\rho, \phi)$ as variable,  we find that the deterministic parts of the resulting field equations are equivalent (up to a transformation of variables) to those of
(\ref{eq:LangevinPhi},\ref{eq:LangevinRho}), but the stochastic terms differ. While $\left< \delta \theta^2 \right> = \frac{2}{N^2 \rho}\left(1-\theta g(\theta)\right)$ as in (\ref{eq:KMTheta2}), $\left<\delta \phi^2 \right> = \frac{2\rho}{N^2}$, with the $1-\theta g(\theta)$ term missing even under a change of variables via an It\^o transformation. This corresponds to an additive noise term in the field equations, rather than the multiplicative voter-like noise that is a feature of Eq.~(\ref{eq:LangevinTheta}).

This difference turns out to be highly significant, as can be seen from Fig.~\ref{fig:OrdParamVsEta}. We find that when we omit the $1-\theta^2$ term in the stochastic prefactor in Eq.~(\ref{eq:LangevinTheta}), which is equivalent to integrating the field equations in $\phi$, the global velocity is negligible for all values of $\eta$. This corresponds to the disordered paramagnetic state as described for the agent-based model above.

The reason why the two sets of field equations differ in their phenomenology lies in truncating the Kramers-Moyal expansion at second order: equivalence would be restored if we were to keep all orders in the expansion. However, we cannot then couch the dynamics in an SDE of the usual form. What appears to be important for the present application is that the noise vanishes at the boundary points $\theta=\pm1$ (or $\phi=\pm\rho$) as $\eta\to0$, which happens at second order when we choose to represent the system using $\theta$, but not when we choose $\phi$.

\subsection{\label{sec:Pulse} Strong low-density fluctuations}
 
A notable feature of the noise in Eq.~(\ref{eq:LangevinTheta}) is that its variance is inversely proportional to the density, $\rho$, indicating that its effect on the polarization is stronger in low-density regions. For a single large flock, the strongest stochastic effects should thus be found in the nose and tail. It was shown in \cite{OLoan1999_AlternatingFlock} that minor fluctuations in the nose of the flock were sufficient to cause the flock to alternate direction. These minor fluctuations are generated either by stochastic alternating in the nose, or by head-on collision with a small flock coming in the opposite direction.

To illustrate the importance of stochastic fluctuations in low density regions, we modified the velocity function, $g(\theta)$, in two different ways. For sites where the number of particles below a certain threshold ($\rho < \frac{1}{500}$), we let $g(\theta) = (1-2\eta)\theta$ to simulate run-and-tumble motion or $g(\theta) = 0$ to simulate unbiased diffusion only. Both of these modifications cause the coherent flock to disappear. This suggests that stochastic effects in low density regions such as the noses and tails of flocks are vital for flocks to maintain their structure and avoid collapse.
 
\begin{figure}
    \centering
    \includegraphics[scale=0.25]{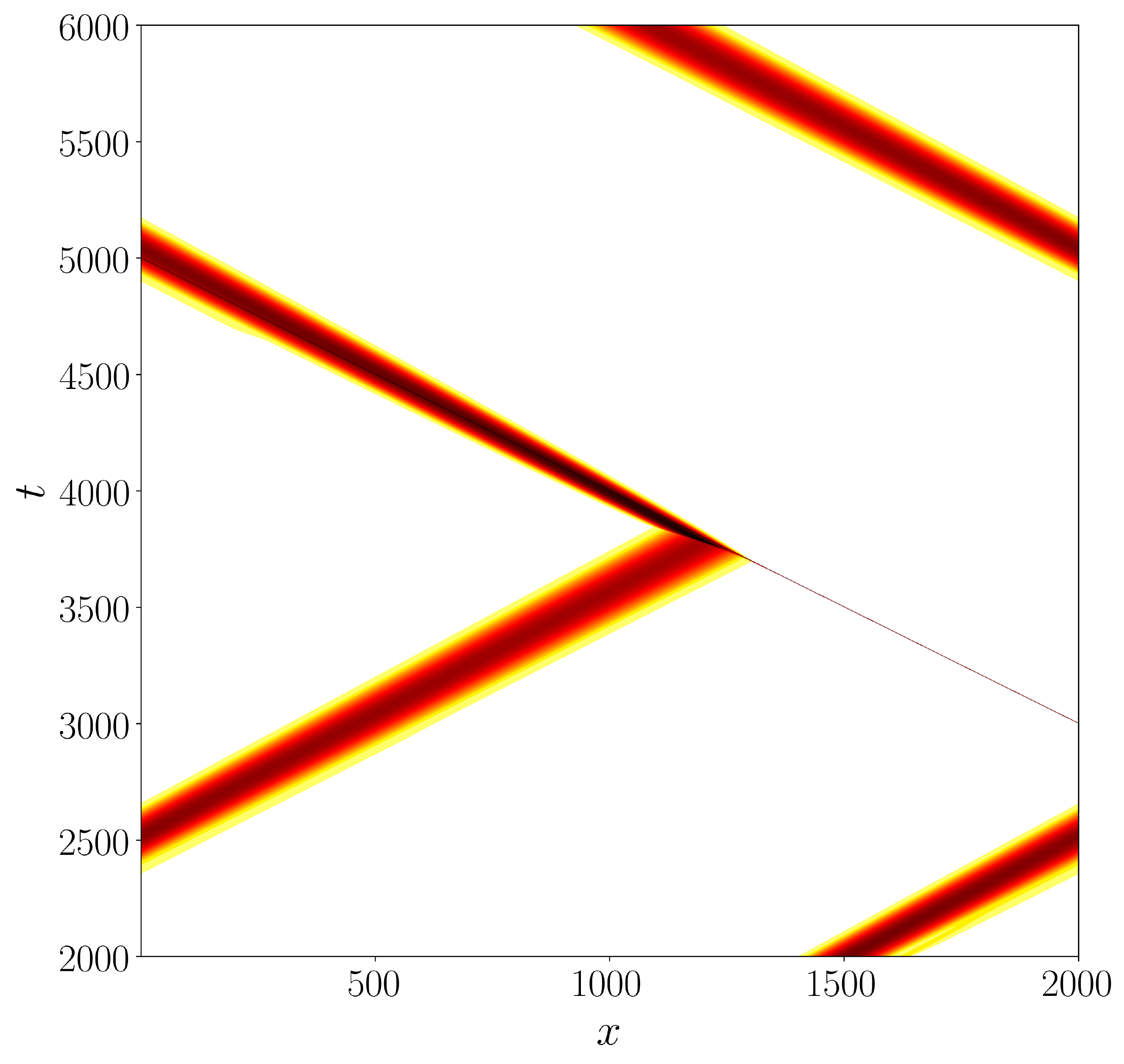}
    \caption{Numerically integrated solution of (\ref{eq:LangevinRho}) with only the deterministic part of (\ref{eq:LangevinTheta}) included.  In the absence of any stochasticity, the flock will diffuse until it has spread over the system. However, the flock alternates direction when a smaller flock impacts the nose. This artificially introduced small flock can be seen as a thin line coming from the right edge at $t=3000$. Here, we have chosen $L=N=2000$, $\eta = 0.02$ and $\beta = 2.0$}
    \label{fig:TurnFlock}
\end{figure}

We also find that, even when stochastic contributions are ignored, a large flock is deterministically unstable to collision with an oncoming smaller flock. We achieved this by setting the stochastic noise strength to zero and observing the effect of sending a small pulse (the equivalent of 3 particles) at the resulting flock. Fig. \ref{fig:TurnFlock} shows this occurring. Numerically integrating (\ref{eq:LangevinRho}, \ref{eq:LangevinTheta}) using Beta noise as before also displays this behaviour, as can be seen in Fig. \ref{fig:ThetaNoise}b.

These findings reveal that the stochastic field equations reproduce in detail the physical mechanisms responsible for the alternating state that is seen in the agent-based model.

\subsection{\label{sec:NonLinear}Necessity of a non-linear alignment interaction}

Having established that the stochastic field equations correspond to the agent-based model for the specific choice of the alignment interaction term $g(\theta)$ given by (\ref{eq:gdef}), we now use them to understand the space of functional forms of $g(\theta)$ that allow the emergence of alternating flocks. In particular, we find that some  degree of non-linearity is required for the coherent flock to exist.

To see this, we modify the agent-based model to allow additional particle moves. We allow particles to move in the same direction they are facing with probability $p(1-\eta)$ and to switch direction with probability $p\eta$. This is equivalent to run-and-tumble motion with a tumble rate of $2p\eta$ \cite{Schnitzer1993_RunAndTumble}. To balance probabilities, the movement based on on-site polarization is modified as $W^{\pm} = \frac{1-p}{2}\left(1 \pm g(\theta)\right)$.

Following the same Kramers-Moyal approach from above, we arrive at a set of equations for how the density and polarization evolve in time. The only change to (\ref{eq:LangevinRho}) and (\ref{eq:LangevinTheta}) is that $g(\theta)$ is replaced with $g_p(\theta) = (1-p)g(\theta) + p(1-2\eta)\theta$. All changes to the system by the introduction of a run-and-tumble interaction have been reduced to the addition of a linear term in the velocity function.

The special case $p=1$ corresponds to non-interacting run-and-tumble particles, which clearly cannot order as there are no interactions. If $g(\theta)$ is linear, we find that this is equivalent to taking $p=1$, and hence that that linear interactions alone aren't enough to observe flocking. Given this, we can see how this can also be inferred from (\ref{eq:LangevinTheta}) and the constraints on $g$. If the magnitude of $g$ is greater than that of $\theta$, the magnitude of $\theta$ will grow. Thus we require $\left| g \right| > \left| \theta \right|$ for small $\theta$ so that a coherent flock can form and $\left| g \right| < 1$ from the definition of $g$. These conditions force $g$ to be non-linear. We also expect that $g(\theta)$ should be a monotonically increasing function of $\theta$ for flocking to occur.

\begin{figure}
	\centering
	\includegraphics[scale=0.25]{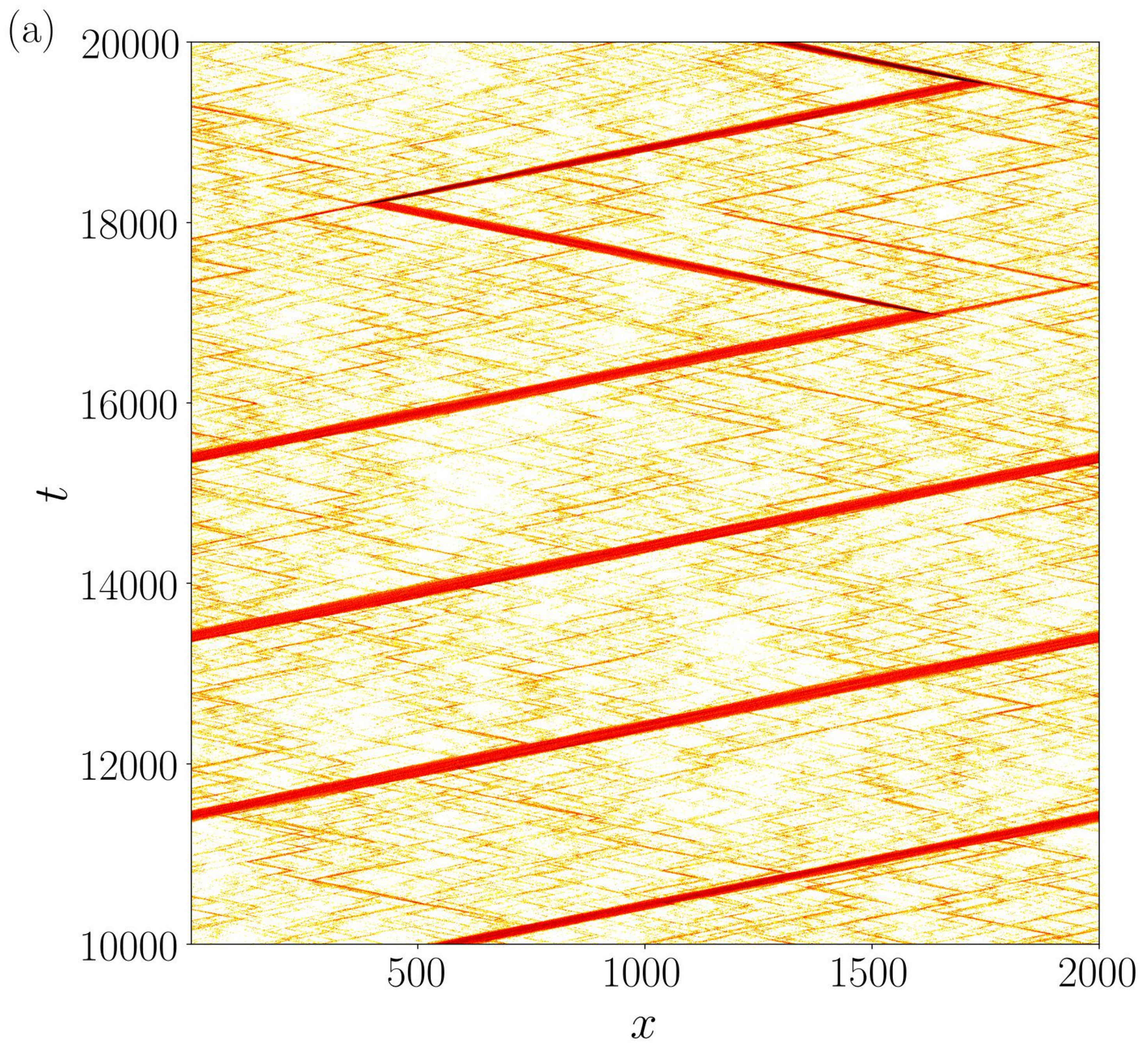}\\
	\includegraphics[scale=0.25]{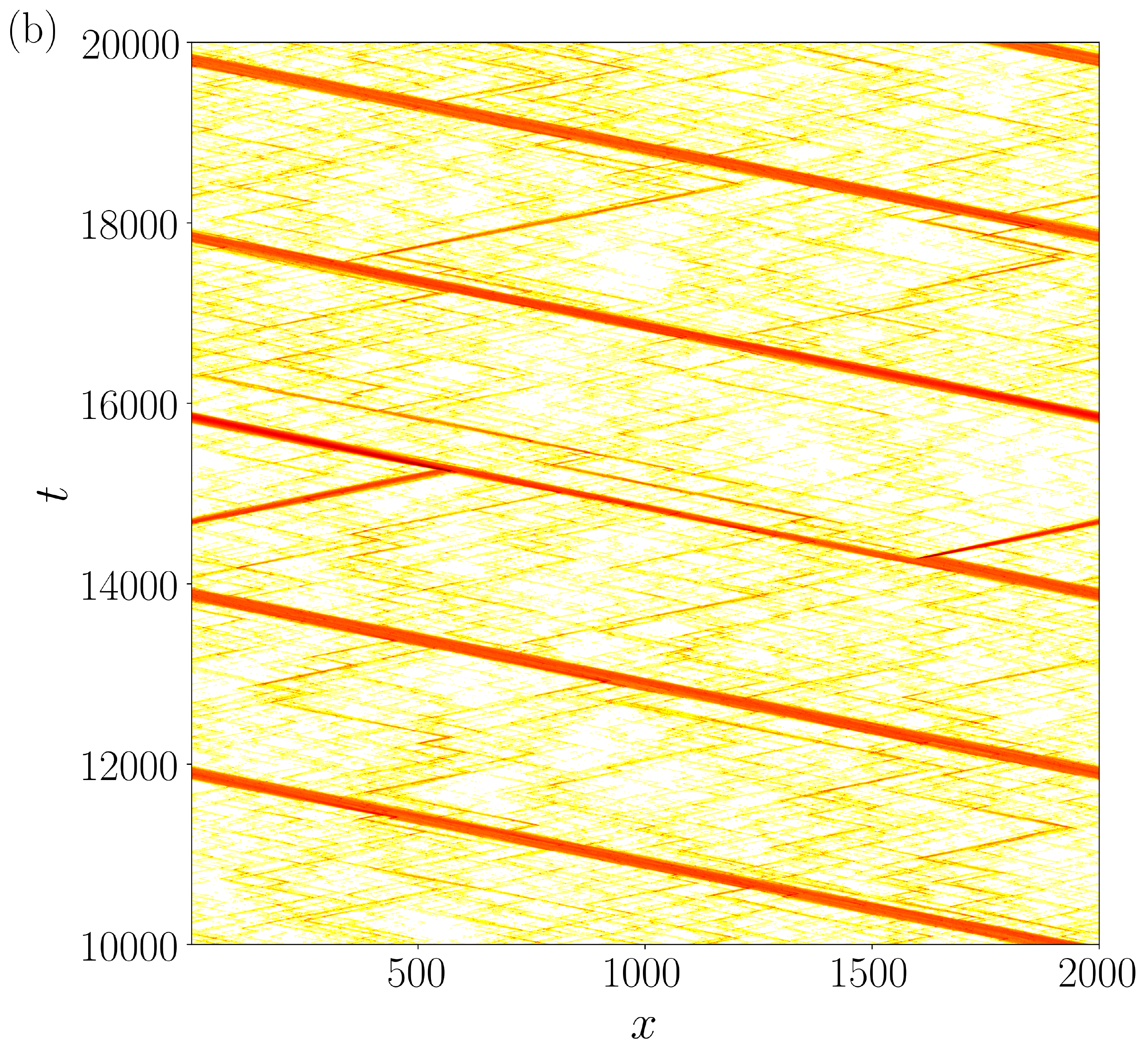}
	\caption{Agent-based model (a) and numerical integration of (\ref{eq:LangevinRho},\ref{eq:LangevinTheta}) (b) with modified $g(\theta)$ as in (\ref{eq:gmin}). Note that both display the same qualitative behaviour as Fig.~\ref{fig:ABM-Phases}b and Fig.~\ref{fig:ThetaNoise}b. Here, we have chosen $L=N=2000$, $\eta = 0.02$ and $\beta = 2.0$}
	\label{fig:MinModel}
\end{figure}

The simplest function that satisfies these requirements is
\begin{equation}
\label{eq:gmin}
g(\theta) = \frac{3(1-2\eta)}{2}\left( x - \frac{x^3}{3} \right) \;,
\end{equation}
which corresponds to keeping the first nonlinear term in the $\tanh$ function in Eq.~(\ref{eq:gdef}). Numerical integration of these equations (see Fig.~\ref{fig:MinModel}) indicates that flocking occurs when $\eta$ is sufficiently small.

Consequently, this leads us to suggest that a minimal model for one-dimensional flocking with an alternating state is given by the stochastic field equations (\ref{eq:LangevinRho}) and (\ref{eq:LangevinTheta}), or equivalently (\ref{eq:LangevinRho}) and (\ref{eq:LangevinPhi}), with $g(\theta)$ approximated by a linear function in the stochastic term, and by a cubic function (\ref{eq:gmin}) in the deterministic terms.

\section{\label{sec:Conclusions}Conclusions}

In this work, we have constructed a set of stochastic field equations, Eqs.~(\ref{eq:LangevinRho}) and (\ref{eq:LangevinTheta}), that exhibited alternating motion without the need for long-range interactions. We found that the precise form of the noise term in the equation for the polarization was crucial to obtain this behaviour. Specifically, we found that the variance of the noise needs to increase as the density decreases (as previously discussed in \cite{Vicsek1999_1dFlocking}) \emph{and} that it must vanish as the polarization approaches its boundary values $\theta=\pm1$. The voter-type noise, with a variance proportional to $(1-\theta^2)/\rho$, is probably the simplest noise with these properties. While the noise term we derive is not precisely the usual voter noise \cite{Dickman1995_VoterTypeNoise,Dornic2001_UniversalityClassVoterModel,RussellBlythe2011_NoiseInducedTransition}, the two multiplicative prefactors agree to first order approximation in $\eta$ and $\theta$. Furthermore, we found that, for flocks to form, a nonlinear alignment interaction is required for the deterministic terms of both density and polarization fields: that is, when a particle adopts a direction to move from its local neighbourhood, it needs to be biased towards the majority direction. Again, a cubic function is the simplest polynomial with these properties, and we have found this sufficient to generate a steady state with an alternating flock. In this sense, we regard the field equations derived here as a minimal model for one-dimensional flocking.

These results highlight the importance of deriving the form of the noise from first principles when constructing continuum models of active particle systems. Whilst at equilibrium, it is usually sufficient to add Gaussian white noise to model the effects of reversible heat exchanges with a reservoir, for internally-driven systems one may miss important aspects of the dynamics. More subtly, we found that the choice of variables used to describe the system (i.e, polarization versus momentum) is significant in determining the form of the noise. This is due to truncating the Kramers-Moyal expansion at second order and the assumption that density and momentum are uncorrelated in space, which are necessary if one wishes to obtain a description that takes the form of a stochastic differential equation. It would be of interest to determine a systematic procedure for identifying the appropriate combination of variables for active particle systems in general, as this appears to be nontrivial.

We also identified a strategy for integrating equations with voter-type noise that gives excellent quantitative agreement between the agent-based model and stochastic field equations. Specifically, instead of using stochastic increments that are drawn from a Gaussian distribution, which is the usual approach to integrating an SDE, we sampled instead from a Beta distribution with the desired mean and variance. The benefit of this approach is that random variables drawn from this distribution are naturally bounded, and therefore one does not need to apply an ad-hoc rule to restrain variables to the physical region. This technique may prove useful in understanding other models with voter-type noise. A further benefit is that one can simulate systems with higher densities with less computational effort than the corresponding agent-based models.

These stochastic field equations have further allowed us to better understand the stability of the alternating state, and to examine the causes of the flock flipping. The inverse density dependence observed in (\ref{eq:LangevinTheta}) suggests that stochastic contributions in regions of low density, such as the nose and tail of the flock, are important, confirming O'Loan \& Evans \cite{OLoan1999_AlternatingFlock}. By modifying the fluctuations in low density regions, we showed that the alternating state was no longer stable. In this case, the large aggregate flock collapsed and the modified behaviour was observed everywhere (run-and-tumble and unbiased random walk were the two examples used). We also showed that the flock was deterministically unstable to changing direction under a head-on collision with a much smaller oncoming flock, also confirming \cite{OLoan1999_AlternatingFlock}. If this smaller flock hadn't existed and in the absence of stochastic contributions, the main flock would have eventually diffused over the full system and the cohesive flock would have been lost.

\section{Acknowledgements}
The authors would like to thank John Toner for useful discussions. E. \'O Laighl\'eis acknowledges studentship funding from EPSRC under grant no. EP/L015110/1.

\bibliography{bibfile}

\begin{thebibliography}{42}%
\makeatletter
\providecommand \@ifxundefined [1]{%
 \@ifx{#1\undefined}
}%
\providecommand \@ifnum [1]{%
 \ifnum #1\expandafter \@firstoftwo
 \else \expandafter \@secondoftwo
 \fi
}%
\providecommand \@ifx [1]{%
 \ifx #1\expandafter \@firstoftwo
 \else \expandafter \@secondoftwo
 \fi
}%
\providecommand \natexlab [1]{#1}%
\providecommand \enquote  [1]{``#1''}%
\providecommand \bibnamefont  [1]{#1}%
\providecommand \bibfnamefont [1]{#1}%
\providecommand \citenamefont [1]{#1}%
\providecommand \href@noop [0]{\@secondoftwo}%
\providecommand \href [0]{\begingroup \@sanitize@url \@href}%
\providecommand \@href[1]{\@@startlink{#1}\@@href}%
\providecommand \@@href[1]{\endgroup#1\@@endlink}%
\providecommand \@sanitize@url [0]{\catcode `\\12\catcode `\$12\catcode
  `\&12\catcode `\#12\catcode `\^12\catcode `\_12\catcode `\%12\relax}%
\providecommand \@@startlink[1]{}%
\providecommand \@@endlink[0]{}%
\providecommand \url  [0]{\begingroup\@sanitize@url \@url }%
\providecommand \@url [1]{\endgroup\@href {#1}{\urlprefix }}%
\providecommand \urlprefix  [0]{URL }%
\providecommand \Eprint [0]{\href }%
\providecommand \doibase [0]{http://dx.doi.org/}%
\providecommand \selectlanguage [0]{\@gobble}%
\providecommand \bibinfo  [0]{\@secondoftwo}%
\providecommand \bibfield  [0]{\@secondoftwo}%
\providecommand \translation [1]{[#1]}%
\providecommand \BibitemOpen [0]{}%
\providecommand \bibitemStop [0]{}%
\providecommand \bibitemNoStop [0]{.\EOS\space}%
\providecommand \EOS [0]{\spacefactor3000\relax}%
\providecommand \BibitemShut  [1]{\csname bibitem#1\endcsname}%
\let\auto@bib@innerbib\@empty
\bibitem [{\citenamefont {Toner}\ \emph {et~al.}(2005)\citenamefont {Toner},
  \citenamefont {Tu},\ and\ \citenamefont
  {Ramaswamy}}]{TonerTu2005_FlockingReview}%
  \BibitemOpen
  \bibfield  {author} {\bibinfo {author} {\bibfnamefont {J.}~\bibnamefont
  {Toner}}, \bibinfo {author} {\bibfnamefont {Y.}~\bibnamefont {Tu}}, \ and\
  \bibinfo {author} {\bibfnamefont {S.}~\bibnamefont {Ramaswamy}},\ }\href
  {https://ac.els-cdn.com/S0003491605000540/1-s2.0-S0003491605000540-main.pdf?{\_}tid=566f5577-1a10-46da-862d-024dd1e03416{\&}acdnat=1527521945{\_}47d6d7f98090a5f9dcc816d21955ec98}
  {\bibfield  {journal} {\bibinfo  {journal} {Ann. Phys. (N. Y).}\ }\textbf
  {\bibinfo {volume} {318}},\ \bibinfo {pages} {170} (\bibinfo {year}
  {2005})}\BibitemShut {NoStop}%
\bibitem [{\citenamefont {Vicsek}\ and\ \citenamefont
  {Zafeiris}(2012)}]{Vicsek2012_CollectiveMotionReview}%
  \BibitemOpen
  \bibfield  {author} {\bibinfo {author} {\bibfnamefont {T.}~\bibnamefont
  {Vicsek}}\ and\ \bibinfo {author} {\bibfnamefont {A.}~\bibnamefont
  {Zafeiris}},\ }\href {\doibase 10.1016/j.physrep.2012.03.004} {\bibfield
  {journal} {\bibinfo  {journal} {Phys. Rep.}\ }\textbf {\bibinfo {volume}
  {517}},\ \bibinfo {pages} {71} (\bibinfo {year} {2012})}\BibitemShut
  {NoStop}%
\bibitem [{\citenamefont {Ramaswamy}(2017)}]{Ramaswamy2017_ActiveMatterReview}%
  \BibitemOpen
  \bibfield  {author} {\bibinfo {author} {\bibfnamefont {S.}~\bibnamefont
  {Ramaswamy}},\ }\href {\doibase 10.1088/1742-5468/aa6bc5} {\bibfield
  {journal} {\bibinfo  {journal} {J. Stat. Mech. Theory Exp.}\ }\textbf
  {\bibinfo {volume} {2017}},\ \bibinfo {pages} {054002} (\bibinfo {year}
  {2017})}\BibitemShut {NoStop}%
\bibitem [{\citenamefont {Pearce}\ \emph {et~al.}(2014)\citenamefont {Pearce},
  \citenamefont {Miller}, \citenamefont {Rowlands},\ and\ \citenamefont
  {Turner}}]{Pearce2014_ProjectionBirdFlocks}%
  \BibitemOpen
  \bibfield  {author} {\bibinfo {author} {\bibfnamefont {D.~J.~G.}\
  \bibnamefont {Pearce}}, \bibinfo {author} {\bibfnamefont {A.~M.}\
  \bibnamefont {Miller}}, \bibinfo {author} {\bibfnamefont {G.}~\bibnamefont
  {Rowlands}}, \ and\ \bibinfo {author} {\bibfnamefont {M.~S.}\ \bibnamefont
  {Turner}},\ }\href {\doibase 10.1073/pnas.1402202111} {\bibfield  {journal}
  {\bibinfo  {journal} {Proc. Natl. Acad. Sci.}\ }\textbf {\bibinfo {volume}
  {111}},\ \bibinfo {pages} {10422} (\bibinfo {year} {2014})}\BibitemShut
  {NoStop}%
\bibitem [{\citenamefont {Lewis}\ and\ \citenamefont
  {Turner}(2017)}]{LewisTurner2017_DensityDepthFlocks}%
  \BibitemOpen
  \bibfield  {author} {\bibinfo {author} {\bibfnamefont {J.~M.}\ \bibnamefont
  {Lewis}}\ and\ \bibinfo {author} {\bibfnamefont {M.~S.}\ \bibnamefont
  {Turner}},\ }\href {\doibase 10.1088/1361-6463/aa942f} {\bibfield  {journal}
  {\bibinfo  {journal} {J. Phys. D. Appl. Phys.}\ }\textbf {\bibinfo {volume}
  {50}},\ \bibinfo {pages} {494003} (\bibinfo {year} {2017})}\BibitemShut
  {NoStop}%
\bibitem [{\citenamefont {Katz}\ \emph {et~al.}(2011)\citenamefont {Katz},
  \citenamefont {Tunstrom}, \citenamefont {Ioannou}, \citenamefont {Huepe},\
  and\ \citenamefont {Couzin}}]{Katz2011_FishFlockingInteractions}%
  \BibitemOpen
  \bibfield  {author} {\bibinfo {author} {\bibfnamefont {Y.}~\bibnamefont
  {Katz}}, \bibinfo {author} {\bibfnamefont {K.}~\bibnamefont {Tunstrom}},
  \bibinfo {author} {\bibfnamefont {C.~C.}\ \bibnamefont {Ioannou}}, \bibinfo
  {author} {\bibfnamefont {C.}~\bibnamefont {Huepe}}, \ and\ \bibinfo {author}
  {\bibfnamefont {I.~D.}\ \bibnamefont {Couzin}},\ }\href {\doibase
  10.1073/pnas.1107583108} {\bibfield  {journal} {\bibinfo  {journal} {Proc.
  Natl. Acad. Sci.}\ }\textbf {\bibinfo {volume} {108}},\ \bibinfo {pages}
  {18720} (\bibinfo {year} {2011})}\BibitemShut {NoStop}%
\bibitem [{\citenamefont {Makris}\ \emph {et~al.}(2009)\citenamefont {Makris},
  \citenamefont {Ratilal}, \citenamefont {Jagannathan}, \citenamefont {Gong},
  \citenamefont {Andrews}, \citenamefont {Bertsatos}, \citenamefont {Godo},
  \citenamefont {Nero},\ and\ \citenamefont
  {Jech}}]{Makris2009_FishExperimentalFlocking}%
  \BibitemOpen
  \bibfield  {author} {\bibinfo {author} {\bibfnamefont {N.~C.}\ \bibnamefont
  {Makris}}, \bibinfo {author} {\bibfnamefont {P.}~\bibnamefont {Ratilal}},
  \bibinfo {author} {\bibfnamefont {S.}~\bibnamefont {Jagannathan}}, \bibinfo
  {author} {\bibfnamefont {Z.}~\bibnamefont {Gong}}, \bibinfo {author}
  {\bibfnamefont {M.}~\bibnamefont {Andrews}}, \bibinfo {author} {\bibfnamefont
  {I.}~\bibnamefont {Bertsatos}}, \bibinfo {author} {\bibfnamefont {O.~R.}\
  \bibnamefont {Godo}}, \bibinfo {author} {\bibfnamefont {R.~W.}\ \bibnamefont
  {Nero}}, \ and\ \bibinfo {author} {\bibfnamefont {J.~M.}\ \bibnamefont
  {Jech}},\ }\href {\doibase 10.1126/science.1169441} {\bibfield  {journal}
  {\bibinfo  {journal} {Science (80-. ).}\ }\textbf {\bibinfo {volume} {323}},\
  \bibinfo {pages} {1734} (\bibinfo {year} {2009})}\BibitemShut {NoStop}%
\bibitem [{\citenamefont {Yates}\ \emph {et~al.}(2009)\citenamefont {Yates},
  \citenamefont {Erban}, \citenamefont {Escudero}, \citenamefont {Couzin},
  \citenamefont {Buhl}, \citenamefont {Kevrekidis}, \citenamefont {Maini},\
  and\ \citenamefont {Sumpter}}]{Yates2009_LocustsAlternatingABM}%
  \BibitemOpen
  \bibfield  {author} {\bibinfo {author} {\bibfnamefont {C.~A.}\ \bibnamefont
  {Yates}}, \bibinfo {author} {\bibfnamefont {R.}~\bibnamefont {Erban}},
  \bibinfo {author} {\bibfnamefont {C.}~\bibnamefont {Escudero}}, \bibinfo
  {author} {\bibfnamefont {I.~D.}\ \bibnamefont {Couzin}}, \bibinfo {author}
  {\bibfnamefont {J.}~\bibnamefont {Buhl}}, \bibinfo {author} {\bibfnamefont
  {I.~G.}\ \bibnamefont {Kevrekidis}}, \bibinfo {author} {\bibfnamefont
  {P.~K.}\ \bibnamefont {Maini}}, \ and\ \bibinfo {author} {\bibfnamefont
  {D.~J.~T.}\ \bibnamefont {Sumpter}},\ }\href {\doibase
  10.1073/pnas.0811195106} {\bibfield  {journal} {\bibinfo  {journal} {Proc.
  Natl. Acad. Sci.}\ }\textbf {\bibinfo {volume} {106}},\ \bibinfo {pages}
  {5464} (\bibinfo {year} {2009})}\BibitemShut {NoStop}%
\bibitem [{\citenamefont {Beekman}\ \emph {et~al.}(2001)\citenamefont
  {Beekman}, \citenamefont {Sumpter},\ and\ \citenamefont
  {Ratnieks}}]{Beekman2001_AntsCollectiveMotionTransition}%
  \BibitemOpen
  \bibfield  {author} {\bibinfo {author} {\bibfnamefont {M.}~\bibnamefont
  {Beekman}}, \bibinfo {author} {\bibfnamefont {D.~J.~T.}\ \bibnamefont
  {Sumpter}}, \ and\ \bibinfo {author} {\bibfnamefont {F.~L.~W.}\ \bibnamefont
  {Ratnieks}},\ }\href {\doibase 10.1073/pnas.161285298} {\bibfield  {journal}
  {\bibinfo  {journal} {Proc. Natl. Acad. Sci.}\ }\textbf {\bibinfo {volume}
  {98}},\ \bibinfo {pages} {9703} (\bibinfo {year} {2001})}\BibitemShut
  {NoStop}%
\bibitem [{\citenamefont {Wu}\ \emph {et~al.}(2009)\citenamefont {Wu},
  \citenamefont {Kaiser}, \citenamefont {Jiang},\ and\ \citenamefont
  {Alber}}]{Wu2009_BacteriaSwarmingAlternatingDirection}%
  \BibitemOpen
  \bibfield  {author} {\bibinfo {author} {\bibfnamefont {Y.}~\bibnamefont
  {Wu}}, \bibinfo {author} {\bibfnamefont {A.~D.}\ \bibnamefont {Kaiser}},
  \bibinfo {author} {\bibfnamefont {Y.}~\bibnamefont {Jiang}}, \ and\ \bibinfo
  {author} {\bibfnamefont {M.~S.}\ \bibnamefont {Alber}},\ }\href {\doibase
  10.1073/pnas.0811662106} {\bibfield  {journal} {\bibinfo  {journal} {Proc.
  Natl. Acad. Sci.}\ }\textbf {\bibinfo {volume} {106}},\ \bibinfo {pages}
  {1222} (\bibinfo {year} {2009})}\BibitemShut {NoStop}%
\bibitem [{\citenamefont {Vicsek}\ \emph {et~al.}(1995)\citenamefont {Vicsek},
  \citenamefont {Czir{\'{o}}k}, \citenamefont {Ben-Jacob}, \citenamefont
  {Cohen},\ and\ \citenamefont {Shochet}}]{Vicsek1995_CollectiveMotionModel}%
  \BibitemOpen
  \bibfield  {author} {\bibinfo {author} {\bibfnamefont {T.}~\bibnamefont
  {Vicsek}}, \bibinfo {author} {\bibfnamefont {A.}~\bibnamefont
  {Czir{\'{o}}k}}, \bibinfo {author} {\bibfnamefont {E.}~\bibnamefont
  {Ben-Jacob}}, \bibinfo {author} {\bibfnamefont {I.}~\bibnamefont {Cohen}}, \
  and\ \bibinfo {author} {\bibfnamefont {O.}~\bibnamefont {Shochet}},\ }\href
  {\doibase 10.1103/PhysRevLett.75.1226} {\bibfield  {journal} {\bibinfo
  {journal} {Phys. Rev. Lett.}\ }\textbf {\bibinfo {volume} {75}},\ \bibinfo
  {pages} {1226} (\bibinfo {year} {1995})}\BibitemShut {NoStop}%
\bibitem [{\citenamefont {Czir{\'{o}}k}\ \emph {et~al.}(1999)\citenamefont
  {Czir{\'{o}}k}, \citenamefont {Barab{\'{a}}si},\ and\ \citenamefont
  {Vicsek}}]{Vicsek1999_1dFlocking}%
  \BibitemOpen
  \bibfield  {author} {\bibinfo {author} {\bibfnamefont {A.}~\bibnamefont
  {Czir{\'{o}}k}}, \bibinfo {author} {\bibfnamefont {A.-L.}\ \bibnamefont
  {Barab{\'{a}}si}}, \ and\ \bibinfo {author} {\bibfnamefont {T.}~\bibnamefont
  {Vicsek}},\ }\href {\doibase 10.1103/PhysRevLett.82.209} {\bibfield
  {journal} {\bibinfo  {journal} {Phys. Rev. Lett.}\ }\textbf {\bibinfo
  {volume} {82}},\ \bibinfo {pages} {209} (\bibinfo {year} {1999})}\BibitemShut
  {NoStop}%
\bibitem [{\citenamefont {Chat{\'{e}}}\ \emph {et~al.}(2008)\citenamefont
  {Chat{\'{e}}}, \citenamefont {Ginelli}, \citenamefont {Gr{\'{e}}goire},\ and\
  \citenamefont {Raynaud}}]{Chate2008_VicsekFirstOrderTransition}%
  \BibitemOpen
  \bibfield  {author} {\bibinfo {author} {\bibfnamefont {H.}~\bibnamefont
  {Chat{\'{e}}}}, \bibinfo {author} {\bibfnamefont {F.}~\bibnamefont
  {Ginelli}}, \bibinfo {author} {\bibfnamefont {G.}~\bibnamefont
  {Gr{\'{e}}goire}}, \ and\ \bibinfo {author} {\bibfnamefont {F.}~\bibnamefont
  {Raynaud}},\ }\href {\doibase 10.1103/PhysRevE.77.046113} {\bibfield
  {journal} {\bibinfo  {journal} {Phys. Rev. E}\ }\textbf {\bibinfo {volume}
  {77}},\ \bibinfo {pages} {046113} (\bibinfo {year} {2008})}\BibitemShut
  {NoStop}%
\bibitem [{\citenamefont {Toner}\ and\ \citenamefont
  {Tu}(1998)}]{TonerTu1998_FieldModelFlocking}%
  \BibitemOpen
  \bibfield  {author} {\bibinfo {author} {\bibfnamefont {J.}~\bibnamefont
  {Toner}}\ and\ \bibinfo {author} {\bibfnamefont {Y.}~\bibnamefont {Tu}},\
  }\href {\doibase 10.1103/PhysRevE.58.4828} {\bibfield  {journal} {\bibinfo
  {journal} {Phys. Rev. E}\ }\textbf {\bibinfo {volume} {58}},\ \bibinfo
  {pages} {4828} (\bibinfo {year} {1998})}\BibitemShut {NoStop}%
\bibitem [{\citenamefont {Solon}\ and\ \citenamefont
  {Tailleur}(2013)}]{SolonTailleur2013_ActiveIsingFlockingTransition}%
  \BibitemOpen
  \bibfield  {author} {\bibinfo {author} {\bibfnamefont {A.~P.}\ \bibnamefont
  {Solon}}\ and\ \bibinfo {author} {\bibfnamefont {J.}~\bibnamefont
  {Tailleur}},\ }\href {\doibase 10.1103/PhysRevLett.111.078101} {\bibfield
  {journal} {\bibinfo  {journal} {Phys. Rev. Lett.}\ }\textbf {\bibinfo
  {volume} {111}},\ \bibinfo {pages} {078101} (\bibinfo {year}
  {2013})}\BibitemShut {NoStop}%
\bibitem [{\citenamefont {O'Loan}\ and\ \citenamefont
  {Evans}(1999)}]{OLoan1999_AlternatingFlock}%
  \BibitemOpen
  \bibfield  {author} {\bibinfo {author} {\bibfnamefont {O.~J.}\ \bibnamefont
  {O'Loan}}\ and\ \bibinfo {author} {\bibfnamefont {M.~R.}\ \bibnamefont
  {Evans}},\ }\href {\doibase 10.1088/0305-4470/32/8/002} {\bibfield  {journal}
  {\bibinfo  {journal} {J. Phys. A. Math. Gen.}\ }\textbf {\bibinfo {volume}
  {32}},\ \bibinfo {pages} {L99} (\bibinfo {year} {1999})}\BibitemShut
  {NoStop}%
\bibitem [{\citenamefont {Czir{\'{o}}k}\ and\ \citenamefont
  {Vicsek}(2000)}]{Vicsek2000_EarlyFlockingReview}%
  \BibitemOpen
  \bibfield  {author} {\bibinfo {author} {\bibfnamefont {A.}~\bibnamefont
  {Czir{\'{o}}k}}\ and\ \bibinfo {author} {\bibfnamefont {T.}~\bibnamefont
  {Vicsek}},\ }\href {\doibase 10.1016/S0378-4371(00)00013-3} {\bibfield
  {journal} {\bibinfo  {journal} {Phys. A Stat. Mech. its Appl.}\ }\textbf
  {\bibinfo {volume} {281}},\ \bibinfo {pages} {17} (\bibinfo {year}
  {2000})}\BibitemShut {NoStop}%
\bibitem [{\citenamefont {Solon}\ \emph
  {et~al.}(2015{\natexlab{a}})\citenamefont {Solon}, \citenamefont
  {Chat{\'{e}}},\ and\ \citenamefont
  {Tailleur}}]{SolonChateTailleur2015_LiquidGasVicsek}%
  \BibitemOpen
  \bibfield  {author} {\bibinfo {author} {\bibfnamefont {A.~P.}\ \bibnamefont
  {Solon}}, \bibinfo {author} {\bibfnamefont {H.}~\bibnamefont {Chat{\'{e}}}},
  \ and\ \bibinfo {author} {\bibfnamefont {J.}~\bibnamefont {Tailleur}},\
  }\href {\doibase 10.1103/PhysRevLett.114.068101} {\bibfield  {journal}
  {\bibinfo  {journal} {Phys. Rev. Lett.}\ }\textbf {\bibinfo {volume} {114}},\
  \bibinfo {pages} {068101} (\bibinfo {year} {2015}{\natexlab{a}})}\BibitemShut
  {NoStop}%
\bibitem [{\citenamefont {Solon}\ \emph
  {et~al.}(2015{\natexlab{b}})\citenamefont {Solon}, \citenamefont {Caussin},
  \citenamefont {Bartolo}, \citenamefont {Chat{\'{e}}},\ and\ \citenamefont
  {Tailleur}}]{Solon2015_IsingVicsekLiquidGas}%
  \BibitemOpen
  \bibfield  {author} {\bibinfo {author} {\bibfnamefont {A.~P.}\ \bibnamefont
  {Solon}}, \bibinfo {author} {\bibfnamefont {J.-B.}\ \bibnamefont {Caussin}},
  \bibinfo {author} {\bibfnamefont {D.}~\bibnamefont {Bartolo}}, \bibinfo
  {author} {\bibfnamefont {H.}~\bibnamefont {Chat{\'{e}}}}, \ and\ \bibinfo
  {author} {\bibfnamefont {J.}~\bibnamefont {Tailleur}},\ }\href {\doibase
  10.1103/PhysRevE.92.062111} {\bibfield  {journal} {\bibinfo  {journal} {Phys.
  Rev. E}\ }\textbf {\bibinfo {volume} {92}},\ \bibinfo {pages} {062111}
  (\bibinfo {year} {2015}{\natexlab{b}})}\BibitemShut {NoStop}%
\bibitem [{\citenamefont {Eftimie}\ \emph {et~al.}(2007)\citenamefont
  {Eftimie}, \citenamefont {de~Vries},\ and\ \citenamefont
  {Lewis}}]{Eftimie2007_FlockingPatterns}%
  \BibitemOpen
  \bibfield  {author} {\bibinfo {author} {\bibfnamefont {R.}~\bibnamefont
  {Eftimie}}, \bibinfo {author} {\bibfnamefont {G.}~\bibnamefont {de~Vries}}, \
  and\ \bibinfo {author} {\bibfnamefont {M.~A.}\ \bibnamefont {Lewis}},\ }\href
  {\doibase 10.1073/pnas.0611483104} {\bibfield  {journal} {\bibinfo  {journal}
  {Proc. Natl. Acad. Sci.}\ }\textbf {\bibinfo {volume} {104}},\ \bibinfo
  {pages} {6974} (\bibinfo {year} {2007})}\BibitemShut {NoStop}%
\bibitem [{\citenamefont {Eftimie}(2013)}]{Eftimie2013_CollectiveBehaviour}%
  \BibitemOpen
  \bibfield  {author} {\bibinfo {author} {\bibfnamefont {R.}~\bibnamefont
  {Eftimie}},\ }\href {\doibase 10.1016/j.jtbi.2013.08.001} {\bibfield
  {journal} {\bibinfo  {journal} {J. Theor. Biol.}\ }\textbf {\bibinfo {volume}
  {337}},\ \bibinfo {pages} {42} (\bibinfo {year} {2013})}\BibitemShut
  {NoStop}%
\bibitem [{\citenamefont {Dyson}\ \emph {et~al.}(2015)\citenamefont {Dyson},
  \citenamefont {Yates}, \citenamefont {Buhl},\ and\ \citenamefont
  {McKane}}]{Dyson2015_LocustFlockingFieldModel}%
  \BibitemOpen
  \bibfield  {author} {\bibinfo {author} {\bibfnamefont {L.}~\bibnamefont
  {Dyson}}, \bibinfo {author} {\bibfnamefont {C.~A.}\ \bibnamefont {Yates}},
  \bibinfo {author} {\bibfnamefont {J.}~\bibnamefont {Buhl}}, \ and\ \bibinfo
  {author} {\bibfnamefont {A.~J.}\ \bibnamefont {McKane}},\ }\href {\doibase
  10.1103/PhysRevE.92.052708} {\bibfield  {journal} {\bibinfo  {journal} {Phys.
  Rev. E}\ }\textbf {\bibinfo {volume} {92}},\ \bibinfo {pages} {052708}
  (\bibinfo {year} {2015})}\BibitemShut {NoStop}%
\bibitem [{\citenamefont {Eftimie}(2012)}]{Eftimie2012_FlockingModelsReview}%
  \BibitemOpen
  \bibfield  {author} {\bibinfo {author} {\bibfnamefont {R.}~\bibnamefont
  {Eftimie}},\ }\href {\doibase 10.1007/s00285-011-0452-2} {\bibfield
  {journal} {\bibinfo  {journal} {J. Math. Biol.}\ }\textbf {\bibinfo {volume}
  {65}},\ \bibinfo {pages} {35} (\bibinfo {year} {2012})}\BibitemShut {NoStop}%
\bibitem [{\citenamefont
  {Gardiner}(2009)}]{Gardiner2009_HandbookStochasticMethods}%
  \BibitemOpen
  \bibfield  {author} {\bibinfo {author} {\bibfnamefont {C.~W.}\ \bibnamefont
  {Gardiner}},\ }\href@noop {} {\emph {\bibinfo {title} {{Stochastic methods: a
  handbook for the natural and social sciences}}}}\ (\bibinfo  {publisher}
  {Springer Berlin},\ \bibinfo {year} {2009})\BibitemShut {NoStop}%
\bibitem [{\citenamefont {Dickman}\ and\ \citenamefont
  {Tretyakov}(1995)}]{Dickman1995_VoterTypeNoise}%
  \BibitemOpen
  \bibfield  {author} {\bibinfo {author} {\bibfnamefont {R.}~\bibnamefont
  {Dickman}}\ and\ \bibinfo {author} {\bibfnamefont {A.~Y.}\ \bibnamefont
  {Tretyakov}},\ }\href {\doibase 10.1103/PhysRevE.52.3218} {\bibfield
  {journal} {\bibinfo  {journal} {Phys. Rev. E}\ }\textbf {\bibinfo {volume}
  {52}},\ \bibinfo {pages} {3218} (\bibinfo {year} {1995})}\BibitemShut
  {NoStop}%
\bibitem [{\citenamefont {Dornic}\ \emph {et~al.}(2001)\citenamefont {Dornic},
  \citenamefont {Chat{\'{e}}}, \citenamefont {Chave},\ and\ \citenamefont
  {Hinrichsen}}]{Dornic2001_UniversalityClassVoterModel}%
  \BibitemOpen
  \bibfield  {author} {\bibinfo {author} {\bibfnamefont {I.}~\bibnamefont
  {Dornic}}, \bibinfo {author} {\bibfnamefont {H.}~\bibnamefont {Chat{\'{e}}}},
  \bibinfo {author} {\bibfnamefont {J.}~\bibnamefont {Chave}}, \ and\ \bibinfo
  {author} {\bibfnamefont {H.}~\bibnamefont {Hinrichsen}},\ }\href {\doibase
  10.1103/PhysRevLett.87.045701} {\bibfield  {journal} {\bibinfo  {journal}
  {Phys. Rev. Lett.}\ }\textbf {\bibinfo {volume} {87}},\ \bibinfo {pages}
  {045701} (\bibinfo {year} {2001})}\BibitemShut {NoStop}%
\bibitem [{\citenamefont {Russell}\ and\ \citenamefont
  {Blythe}(2011)}]{RussellBlythe2011_NoiseInducedTransition}%
  \BibitemOpen
  \bibfield  {author} {\bibinfo {author} {\bibfnamefont {D.~I.}\ \bibnamefont
  {Russell}}\ and\ \bibinfo {author} {\bibfnamefont {R.~A.}\ \bibnamefont
  {Blythe}},\ }\href {\doibase 10.1103/PhysRevLett.106.165702} {\bibfield
  {journal} {\bibinfo  {journal} {Phys. Rev. Lett.}\ }\textbf {\bibinfo
  {volume} {106}},\ \bibinfo {pages} {165702} (\bibinfo {year}
  {2011})}\BibitemShut {NoStop}%
\bibitem [{\citenamefont {Schnitzer}(1993)}]{Schnitzer1993_RunAndTumble}%
  \BibitemOpen
  \bibfield  {author} {\bibinfo {author} {\bibfnamefont {M.~J.}\ \bibnamefont
  {Schnitzer}},\ }\href {\doibase 10.1103/PhysRevE.48.2553} {\bibfield
  {journal} {\bibinfo  {journal} {Phys. Rev. E}\ }\textbf {\bibinfo {volume}
  {48}},\ \bibinfo {pages} {2553} (\bibinfo {year} {1993})}\BibitemShut
  {NoStop}%
\bibitem [{\citenamefont {Ramaswamy}(2010)}]{Ramaswamy2010_ActiveMatterReview}%
  \BibitemOpen
  \bibfield  {author} {\bibinfo {author} {\bibfnamefont {S.}~\bibnamefont
  {Ramaswamy}},\ }\href {\doibase 10.1146/annurev-conmatphys-070909-104101}
  {\bibfield  {journal} {\bibinfo  {journal} {Annu. Rev. Condens. Matter
  Phys.}\ }\textbf {\bibinfo {volume} {1}},\ \bibinfo {pages} {323} (\bibinfo
  {year} {2010})}\BibitemShut {NoStop}%
\bibitem [{\citenamefont {Raymond}\ and\ \citenamefont
  {Evans}(2006)}]{Raymond2006_Flocking}%
  \BibitemOpen
  \bibfield  {author} {\bibinfo {author} {\bibfnamefont {J.~R.}\ \bibnamefont
  {Raymond}}\ and\ \bibinfo {author} {\bibfnamefont {M.~R.}\ \bibnamefont
  {Evans}},\ }\href {\doibase 10.1103/PhysRevE.73.036112} {\bibfield  {journal}
  {\bibinfo  {journal} {Phys. Rev. E}\ }\textbf {\bibinfo {volume} {73}},\
  \bibinfo {pages} {036112} (\bibinfo {year} {2006})}\BibitemShut {NoStop}%
\bibitem [{\citenamefont {Kramers}(1940)}]{Kramers1940_FPDerivation}%
  \BibitemOpen
  \bibfield  {author} {\bibinfo {author} {\bibfnamefont {H.}~\bibnamefont
  {Kramers}},\ }\href {\doibase 10.1016/S0031-8914(40)90098-2} {\bibfield
  {journal} {\bibinfo  {journal} {Physica}\ }\textbf {\bibinfo {volume} {7}},\
  \bibinfo {pages} {284} (\bibinfo {year} {1940})}\BibitemShut {NoStop}%
\bibitem [{\citenamefont {Moyal}(1949)}]{Moyal1949_FPDerivation}%
  \BibitemOpen
  \bibfield  {author} {\bibinfo {author} {\bibfnamefont {J.}~\bibnamefont
  {Moyal}},\ }\href
  {https://www.jstor.org/stable/pdf/2984076.pdf?refreqid=excelsior{\%}3A637cfd08823912f83ad046cd06beeb95}
  {\bibfield  {journal} {\bibinfo  {journal} {J. R. Stat. Soc. Ser. B}\
  }\textbf {\bibinfo {volume} {11}},\ \bibinfo {pages} {150} (\bibinfo {year}
  {1949})}\BibitemShut {NoStop}%
\bibitem [{\citenamefont
  {Pavliotis}(2014)}]{Pavliotis2014_StochasticProcesses}%
  \BibitemOpen
  \bibfield  {author} {\bibinfo {author} {\bibfnamefont {G.~A.}\ \bibnamefont
  {Pavliotis}},\ }\href {\doibase 10.1007/978-1-4939-1323-7} {\emph {\bibinfo
  {title} {Book}}},\ \bibinfo {series} {Texts in Applied Mathematics},
  Vol.~\bibinfo {volume} {60}\ (\bibinfo  {publisher} {Springer New York},\
  \bibinfo {address} {New York, NY},\ \bibinfo {year} {2014})\ pp.\ \bibinfo
  {pages} {1--345}\BibitemShut {NoStop}%
\bibitem [{\citenamefont {Dereniowski}\ and\ \citenamefont
  {Kubale}(2004)}]{Dereniowski2004_CholeskyDecomposition}%
  \BibitemOpen
  \bibfield  {author} {\bibinfo {author} {\bibfnamefont {D.}~\bibnamefont
  {Dereniowski}}\ and\ \bibinfo {author} {\bibfnamefont {M.}~\bibnamefont
  {Kubale}},\ }\href {\doibase 10.1007/978-3-540-24669-5_127} {\emph {\bibinfo
  {title} {{Cholesky Factorization of Matrices in Parallel and Ranking of
  Graphs}}}}\ (\bibinfo  {publisher} {Springer, Berlin, Heidelberg},\ \bibinfo
  {year} {2004})\ pp.\ \bibinfo {pages} {985--992}\BibitemShut {NoStop}%
\bibitem [{\citenamefont {Bertin}\ \emph {et~al.}(2013)\citenamefont {Bertin},
  \citenamefont {Chat{\'{e}}}, \citenamefont {Ginelli}, \citenamefont {Mishra},
  \citenamefont {Peshkov},\ and\ \citenamefont
  {Ramaswamy}}]{Bertin2013_ActiveNematics}%
  \BibitemOpen
  \bibfield  {author} {\bibinfo {author} {\bibfnamefont {E.}~\bibnamefont
  {Bertin}}, \bibinfo {author} {\bibfnamefont {H.}~\bibnamefont {Chat{\'{e}}}},
  \bibinfo {author} {\bibfnamefont {F.}~\bibnamefont {Ginelli}}, \bibinfo
  {author} {\bibfnamefont {S.}~\bibnamefont {Mishra}}, \bibinfo {author}
  {\bibfnamefont {A.}~\bibnamefont {Peshkov}}, \ and\ \bibinfo {author}
  {\bibfnamefont {S.}~\bibnamefont {Ramaswamy}},\ }\href {\doibase
  10.1088/1367-2630/15/8/085032} {\bibfield  {journal} {\bibinfo  {journal}
  {New J. Phys.}\ }\textbf {\bibinfo {volume} {15}},\ \bibinfo {pages} {085032}
  (\bibinfo {year} {2013})}\BibitemShut {NoStop}%
\bibitem [{\citenamefont {Dean}(1996)}]{Dean1996_ConservedNoise}%
  \BibitemOpen
  \bibfield  {author} {\bibinfo {author} {\bibfnamefont {D.~S.}\ \bibnamefont
  {Dean}},\ }\href {\doibase 10.1088/0305-4470/29/24/001} {\bibfield  {journal}
  {\bibinfo  {journal} {J. Phys. A. Math. Gen.}\ }\textbf {\bibinfo {volume}
  {29}},\ \bibinfo {pages} {L613} (\bibinfo {year} {1996})}\BibitemShut
  {NoStop}%
\bibitem [{\citenamefont {Crow}\ and\ \citenamefont
  {Kimura}(1970)}]{Crow1970_WrightFisher}%
  \BibitemOpen
  \bibfield  {author} {\bibinfo {author} {\bibfnamefont {J.~F.}\ \bibnamefont
  {Crow}}\ and\ \bibinfo {author} {\bibfnamefont {M.}~\bibnamefont {Kimura}},\
  }\href {\doibase 10.2307/1529706} {\emph {\bibinfo {title} {{An Introduction
  to Population Genetics Theory}}}}\ (\bibinfo  {publisher} {New York, Evanston
  and London: Harper {\&} Row, Publishers},\ \bibinfo {year}
  {1970})\BibitemShut {NoStop}%
\bibitem [{\citenamefont {Baxter}\ \emph {et~al.}(2007)\citenamefont {Baxter},
  \citenamefont {Blythe},\ and\ \citenamefont
  {McKane}}]{Blythe2007_ExactSolnKolmogorovGeneticDrift}%
  \BibitemOpen
  \bibfield  {author} {\bibinfo {author} {\bibfnamefont {G.}~\bibnamefont
  {Baxter}}, \bibinfo {author} {\bibfnamefont {R.}~\bibnamefont {Blythe}}, \
  and\ \bibinfo {author} {\bibfnamefont {A.}~\bibnamefont {McKane}},\ }\href
  {\doibase 10.1016/j.mbs.2007.01.001} {\bibfield  {journal} {\bibinfo
  {journal} {Math. Biosci.}\ }\textbf {\bibinfo {volume} {209}},\ \bibinfo
  {pages} {124} (\bibinfo {year} {2007})}\BibitemShut {NoStop}%
\bibitem [{\citenamefont
  {Michaud}(2017)}]{Michaud2017_WeakNoiseKramersMoyalModification}%
  \BibitemOpen
  \bibfield  {author} {\bibinfo {author} {\bibfnamefont {J.}~\bibnamefont
  {Michaud}},\ }\href {\doibase 10.1103/PhysRevE.95.022308} {\bibfield
  {journal} {\bibinfo  {journal} {Phys. Rev. E}\ }\textbf {\bibinfo {volume}
  {95}},\ \bibinfo {pages} {022308} (\bibinfo {year} {2017})}\BibitemShut
  {NoStop}%
\bibitem [{\citenamefont {Moro}\ and\ \citenamefont
  {Schurz}(2007)}]{MoroSchurz2007_NIOperatorSplitting}%
  \BibitemOpen
  \bibfield  {author} {\bibinfo {author} {\bibfnamefont {E.}~\bibnamefont
  {Moro}}\ and\ \bibinfo {author} {\bibfnamefont {H.}~\bibnamefont {Schurz}},\
  }\href {\doibase 10.1137/05063725X} {\bibfield  {journal} {\bibinfo
  {journal} {SIAM J. Sci. Comput.}\ }\textbf {\bibinfo {volume} {29}},\
  \bibinfo {pages} {1525} (\bibinfo {year} {2007})}\BibitemShut {NoStop}%
\bibitem [{\citenamefont {Kimura}(1955)}]{Kimura1955_WrightFisherFPSolution}%
  \BibitemOpen
  \bibfield  {author} {\bibinfo {author} {\bibfnamefont {M.}~\bibnamefont
  {Kimura}},\ }\href {\doibase 10.1073/pnas.41.3.144} {\bibfield  {journal}
  {\bibinfo  {journal} {Proc. Natl. Acad. Sci.}\ }\textbf {\bibinfo {volume}
  {41}},\ \bibinfo {pages} {144} (\bibinfo {year} {1955})}\BibitemShut
  {NoStop}%
\bibitem [{\citenamefont {Dornic}\ \emph {et~al.}(2005)\citenamefont {Dornic},
  \citenamefont {Chat{\'{e}}},\ and\ \citenamefont
  {Mu{\~{n}}oz}}]{Dornic2005_MultiplicativeNoise}%
  \BibitemOpen
  \bibfield  {author} {\bibinfo {author} {\bibfnamefont {I.}~\bibnamefont
  {Dornic}}, \bibinfo {author} {\bibfnamefont {H.}~\bibnamefont {Chat{\'{e}}}},
  \ and\ \bibinfo {author} {\bibfnamefont {M.~A.}\ \bibnamefont
  {Mu{\~{n}}oz}},\ }\href {\doibase 10.1103/PhysRevLett.94.100601} {\bibfield
  {journal} {\bibinfo  {journal} {Phys. Rev. Lett.}\ }\textbf {\bibinfo
  {volume} {94}},\ \bibinfo {pages} {100601} (\bibinfo {year}
  {2005})}\BibitemShut {NoStop}%
\end{thebibliography}%
\end{document}